\documentclass[sigconf, nonacm]{acmart}

\usepackage{makecell}
\usepackage{weiwAlgorithm}

\usepackage{graphicx}
\usepackage{balance}  
\usepackage{lipsum}
\usepackage{amsmath}
\usepackage{physics}
\usepackage{subfigure}
\usepackage{tabularx}
\usepackage{color}
\usepackage{colortbl}
\usepackage{float}
\usepackage{listings}
\usepackage{multirow}
\usepackage[normalem]{ulem}
 \usepackage{longtable}
\usepackage{enumitem}
\usepackage{multirow}
\usepackage{booktabs}
\usepackage{capt-of}
\usepackage{pifont}
\usepackage{ulem}
\usepackage{soul}
\usepackage{cancel}
\usepackage{bm}
\usepackage{url}
\usepackage{caption}

\def\BibTeX{{\rm B\kern-.05em{\sc i\kern-.025em b}\kern-.08em
    T\kern-.1667em\lower.7ex\hbox{E}\kern-.125emX}}

 \graphicspath{{./Graph/}, {./Fig/}, {./Legend/}}

\long\def\comment#1{}

\newcounter{example}[section]
\renewcommand{\theexample}{\nthesection.\arabic{example}}
\newenvironment{example}{
     \refstepcounter{example}
     {\vspace{1ex} \noindent\bf  Example  \theexample:}}{
     \vspace{1ex}} 

\newcounter{definition}[section]
\renewcommand{\thedefinition}{\nthesection.\arabic{definition}}

\newcounter{theorem}[section]
\renewcommand{\thetheorem}{\nthesection.\arabic{theorem}}
\newenvironment{theorem}{\begin{em}
        \refstepcounter{theorem}
        {\vspace{1ex} \noindent\bf  Theorem  \thetheorem:}}{
        \end{em}\vspace{1ex}} 

\newcounter{lemma}[section]
\renewcommand{\thelemma}{\nthesection.\arabic{lemma}}

\newcounter{corollary}[section]
\renewcommand{\thecorollary}{\nthesection.\arabic{corollary}}

\newcounter{proposition}[section]
\renewcommand{\theproposition}{\nthesection.\arabic{proposition}}

\newcounter{remark}[section]
\renewcommand{\theremark}{\nthesection.\arabic{remark}}

\newcommand{\proofsketch}{\noindent{\bf Proof Sketch: }}

\newcommand{\nthesection}{\arabic{section}}

\newcommand{\eop}{\hspace*{\fill}\mbox{$\Box$}\vspace*{1ex}}

\newcommand{\stitle}[1]{\vspace{1mm} \noindent{\bf #1}}

\newcommand{\green}[1]{\textcolor{green}{}}

\newcommand{\kw}[1]{{\ensuremath {\mathsf{#1}}}\xspace}

\newcommand\bigutimes{\mathop{\ooalign{$\bigcup$\cr%
   \hfil\raise0.36ex\hbox{$\scriptscriptstyle\boldsymbol{\times}$}\hfil\cr}}}
\newcommand\bigumius{\mathop{\ooalign{$\bigcup$\cr%
   \hfil\raise0.36ex\hbox{$\scriptscriptstyle\boldsymbol{-}$}\hfil\cr}}}

\newcommand{\SIFT}{\kw{SIFT}}
\newcommand{\GIST}{\kw{GIST}}

\newcommand{\Redcaps}{\kw{Redcaps}}
\newcommand{\arXiv}{\kw{arXiv}}

\newcommand{\Surf}{\kw{SURF}}
\newcommand{\VectorMaton}{\kw{VectorMaton}}
\newcommand{\Milvus}{\kw{Milvus}}

\newcommand{\VBASE}{\kw{VBASE}}
\newcommand{\ACORN}{\kw{ACORN}}

\begin{document}

\title{SURF: Substring Filtered Approximate Nearest Neighbor Search}

\author{Jiadong Xie}
\affiliation{%
  \institution{The Chinese University of Hong Kong}
  \country{}
}
\email{jdxie@se.cuhk.edu.hk}

\author{Hao Jiang}
\affiliation{%
  \institution{Xidian University}
  \country{}
}
\email{haojiang0310@stu.xidian.edu.cn}

\author{Yingfan Liu}
\affiliation{%
  \institution{Xidian University}
  \country{}
}
\email{liuyingfan@xidian.edu.cn}

\author{Jeffrey Xu Yu}
\affiliation{%
  \institution{The Hong Kong University of Science and Technology (Guangzhou)}
  \country{}
}
\email{jeffreyxuyu@hkust-gz.edu.cn}

\author{Hui Li}
\affiliation{%
  \institution{Xidian University}
  \country{}
}
\email{hli@xidian.edu.cn}

\author{Jiangtao Cui}
\affiliation{%
  \institution{Xi'an University of Posts and Telecommunications}
  \institution{Xidian University}
  \country{}
}
\email{cuijt@xupt.edu.cn}

\begin{abstract}
Given a collection of objects $\mathcal{O}$ with high-dimensional vectors and associated strings, such as product images and their textual descriptions, filtered approximate nearest neighbor (ANN) search retrieves vectors similar to a query vector $q$ while satisfying a user-specified predicate. We study substring filtered ANN search, where eligible objects are those whose string attributes contain a query string as a substring. Existing methods are sensitive to query selectivity: they either enumerate large qualifying sets or traverse many non-qualifying objects, incurring redundant distance computations. Although \VectorMaton introduces a dedicated index for this problem, its fine-grained automaton states and heuristic index reuse strategy without global optimality guarantees lead to substantial indexing overhead. We propose \Surf, an index featuring a provably minimal suffix automaton that confines graph traversal to objects satisfying the substring predicate. To further reduce its index cost, we develop \kw{ComSURF}, which introduces path and ancestor compression and optimally combines them through a dynamic programming algorithm. We also extend our approach to support SQL \kw{LIKE} predicate queries with two wildcards. We conduct extensive performance studies and confirm the accuracy and efficiency of our approach in comparison with the state-of-the-art approaches.
\end{abstract}


\maketitle

\section{Introduction}

Recent advancements in embedding models utilizing machine learning techniques have revolutionized the representation of various types of data, including images~\cite{nasrabadi1988image}, texts~\cite{MikolovSCCD13}, and audio~\cite{msong}, by transforming them into high-dimensional vectors that capture their semantic information. This has led to a surge in the need for managing high-dimensional data~\cite{abs-2501-11216,PanWL24,AziziEP23,abs-2504-10326,ZhangXCSXCCH00Y23}. When considering a set of objects $\mathcal{O}$ consisting of $n$ objects, each object can be denoted as a $d$-dimensional vector in $\mathbb{R}^d$. 
A fundamental task is to identify similar objects in $\mathcal{O}$ to a given object represented as a vector $v_q\in \mathbb{R}^d$. This task is formalized as approximate nearest neighbor (ANN) search in high-dimensional data management, which aims to find $k$ vectors in $\mathcal{O}$ that are sufficiently close to a given query vector $v_q$. 
In many real-world applications, however, objects are associated not only with vectors but also with string attributes. For example, product images are accompanied by textual descriptions, scientific papers contain titles, authors, and publication venues, and audio files include creator and album metadata. Users therefore often combine vector similarity with string predicates to restrict the search to eligible objects. For instance, Google Multisearch combines string conditions with similarity search over image embeddings~\cite{google_multisearch_2022}. 
Moreover, the combination of semantic similarity and exact string matching appears in a wide range of applications. In agentic, log, and code search, vector retrieval identifies semantically related objects, whereas lexical or grep-style matching remains effective for exact identifiers, paths, and error messages~\cite{GaoDC21,abs-2605-15184,abs-2605-05242}.
Similarly, recent DNA foundation models provide vector embeddings for genomic sequences~\cite{DallaTorre2025NucleotideTransformer},
while exact string matching remains a fundamental operation for locating sequence fragments~\cite{PhoophakdeeZ07}.
Among string constraints, substring filtering is one of the most fundamental and commonly used operations~\cite{ZhanN025}.
Therefore, in this paper, we study the problem of \emph{substring filtered ANN search}. Formally, given a query $q=(v_q,s_q)$, substring filtered ANN search aims to find the approximate nearest neighbors of $v_q$ among the objects in $\mathcal{O}$ whose text attributes contain $s_q$ as a substring.

However, efficiently supporting substring filtered ANN search remains challenging.
Pre-filtering~\cite{WangYGJXLWGLXYY21,WeiWWLZ0C20} enumerates the qualifying objects before vector search, which becomes expensive when the predicate selectivity is high. In contrast, post-filtering~\cite{WangYGJXLWGLXYY21,WeiWWLZ0C20,ZhanSWPLWCLPZC19} and general predicate-aware ANN methods~\cite{ZhangXCSXCCH00Y23,PatelKGZ24} may traverse many non-qualifying objects, particularly at low or moderate selectivities. Similarly, dense-sparse hybrid search~\cite{ChenZCXZWHYLWLY24} combines lexical and semantic similarities but cannot guarantee that all visited objects satisfy the exact substring predicate. Consequently, these approaches either incur substantial costs in enumerating qualifying objects or waste expensive distance computations on non-qualifying objects, resulting in unstable query performance across different selectivity regimes.
The most closely related work is \VectorMaton~\cite{abs-2603-01525}, which introduces an enhanced suffix automaton to organize substring-related vector sets and employs index reuse to reduce indexing costs. However, it has two issues. First, \VectorMaton groups substrings according to their occurrence-position lists, which are more fine-grained than necessary for substring filtered ANN search. Since our problem only requires the set of objects containing a query substring, different occurrence-position lists may correspond to identical object sets and equivalent future query behavior. Treating them as separate states introduces redundant automaton states and index construction. Second, its index reuse strategy relies on greedy, state-wise decisions without global optimality guarantees, and it does not fully leverage proximity-graph properties; instead, it constructs two separate indices for a reused state. Such local decisions result in unnecessary index construction and additional search and result-merging overhead.

To address these issues, we propose a new index structure for substring filtered ANN search, featuring a provably minimal automaton and a globally optimal compression strategy. The main contributions of this work are summarized as follows.
\ding{202} We propose \Surf, a dedicated index for substring-filtered ANN queries, with a key component \kw{MultiSA}, a minimized suffix automaton that groups substrings according to their qualifying object sets under all possible continuations. We prove that \kw{MultiSA} is the minimal deterministic state-output automaton for our problem, thereby eliminating redundant states in the enhanced suffix automaton of \VectorMaton.
\ding{203} To reduce the index cost, we develop \kw{ComSURF}, a compressed variant of \Surf that better exploits the structural properties of proximity graphs. Specifically, we introduce two complementary compression rules, formulate their joint optimization as an exact dynamic programming problem, and develop an efficient algorithm to obtain the globally optimal compression plan.
\ding{204} To efficiently process queries over \kw{ComSURF}, we develop a search algorithm that adaptively combines multiple search strategies. 
We further extend our approach to support SQL \kw{LIKE} predicates containing both single-character (\_) and multi-character (\%) wildcards. 
\ding{205} We conduct extensive experiments on 8 real-world datasets. The results demonstrate that \kw{ComSURF} consistently outperforms state-of-the-art baselines across different query selectivities. Compared with \VectorMaton, \kw{ComSURF} achieves up to 1.99$\times$ higher QPS at comparable recall, while reducing index time by 2.28--5.70$\times$ and index size by 1.78--3.42$\times$. Moreover, it achieves up to two to three orders of magnitude higher QPS than baselines on \kw{LIKE} predicates.

The paper is organized as follows. We present the preliminaries in Section~\ref{sec:pre} and discuss the limitations of existing methods in Section~\ref{sec:existing}. We then provide an overview of our index design in Section~\ref{sec:overview}, present the optimal compression strategy in Section~\ref{sec:optimal-compress}, describe the index construction details in Section~\ref{sec:compressed}, and introduce our search algorithms in Section~\ref{sec:comsurf-search}. Finally, we conduct extensive experiments to demonstrate the effectiveness and efficiency of our approach in Section~\ref{sec:exp}, and conclude the paper in Section~\ref{sec:con}.

\section{Preliminaries}
\label{sec:pre}

\label{sec:problem}
We first formulate the problem studied in this paper.

Let $\mathcal{O}=\{o_1,o_2,\cdots,o_n\}=\{(v_1,s_1),(v_2,s_2),\cdots,(v_n,s_n)\}$ be a dataset consisting of $n$ objects, where each object $o_i$ consists of a $d$-dimensional vector $v_i\in \mathbb{R}^d$, and a string attribute $s_i$. The distance between two vectors, $v_i$ and $v_j$, denoted as
$\delta(v_i,v_j)$, is Euclidean distance in the $L_2$ norm.
Let $\Sigma$ be an alphabet, each object's attribute $s_i$ comprises characters from $\Sigma$, i.e., $s_i \in \Sigma^*$. 
For a string $T=t_1\ldots t_{|T|}$, we denote a substring $t_l\ldots t_{r}$ of $T$ as $T[l,r]$.

\noindent\textbf{Substring Filtered $k$-ANN Query:}
For a query string $s_q$ over $\Sigma$ ($s_q\in \Sigma^*$), let $\mathcal{O}_{s_q}\subseteq \mathcal{O}$ be the set of objects for which $s_q$ is a substring of their attributes, i.e., $\mathcal{O}_{s_q}=\{o_i=(v_i,s_i)\in \mathcal{O}\mid s_q$ is a substring of $s_i\}$.
The substring filtered $k$-approximate nearest neighbor (SFANN) query $q=(v_q,s_q)$ is to find the $k$-approximate nearest neighbors ($k$-ANN) in $\mathcal{O}_{s_q}$ for user-given query vector $v_q$.

Let $Sel(s_q)$ be the fraction of objects that satisfy the substring filter, i.e., $Sel(s_q)=|\mathcal{O}_{s_q}|/|\mathcal{O}|$. We use $recall@k$ to measure the quality of a $k$-ANN query $q$, which is defined as $|\mathcal{U}\cap \mathcal{R}|/k$, where $\mathcal{U}$ is the ground-truth object set which satisfies the substring filter $s_q$, and $\mathcal{R}$ is the set found by an SFANN method for query $q$.

We then review the preliminary knowledge of the proximity graphs designed for $k$-ANN queries.

\noindent\textbf{Proximity Graphs:}
For a $k$-ANN query $q=(v_q,\emptyset)$, i.e., which aims to find the $k$-ANN from $\mathcal{O}$, the state-of-the-art approaches, e.g., NSG~\cite{nsg} and HNSW~\cite{hnsw}, can support it using a proximity graphs built. Here, a proximity graph $G=(V,E)$, where $v_i\in V$ represents a unique vector $v_i$ (e.g., $v_i$ in $o_i=(v_i,s_i)\in \mathcal{O}$), and an edge $(v_i,v_j)\in E$ represents that $v_j$ is regarded as a close neighbor of $v_i$ under the distance function $\delta(v_i,v_j)$, e.g., $v_j$ may be one of the $k$-nearest neighbors of $v_i$.
Over a PG graph, $G$, constructed, a widely used greedy procedure is beam search~\cite{hnsw,nsg,taumg,fanng,ALMG,dpg,PanWL24,wang2021comprehensive}.
At a high level, beam search maintains a bounded priority queue of promising nodes ordered by their distances to $v_q$. It repeatedly expands the nearest unvisited node, inserts its unseen graph neighbors into the queue, and removes the farthest candidates whenever the queue exceeds the beam width $w$. The process terminates when no retained candidate remains to be expanded, after which the $k$ closest nodes are returned.
Each visited node requires a distance computation against $v_q$, and these computations dominate graph-search latency in practice. Thus, the number of traversed nodes is a central cost measure in our discussion of filtered ANN search.

\label{sec:pre-sa}

We next introduce the suffix automaton for representations of a single string as follows.

\noindent\textbf{Suffix Automaton:}
Suffix automaton (SA)~\cite{BlumerBEHM83,BlumerBHECS85} is a partial deterministic finite automaton that recognizes all suffixes of a given string.
An SA can be viewed as a directed acyclic graph (DAG) $\mathcal{G}=(\mathcal{V},\mathcal{E})$. Here, $\mathcal{V}=\{q[0],q[1],\ldots,q[|\mathcal{V}|-1]\}$ represents a finite set of states, $q[0]$ (also denoted by $q_0$) is the unique source node in $\mathcal{G}$, $F\subseteq \mathcal{V}$ is the set of accepting states,
and $\mathcal{E}$ encapsulates the transitions in $\mathcal{V}\times\Sigma\rightarrow \mathcal{V}$, where $\Sigma$ is an alphabet. Specifically, an edge from $q[i]$ to $q[j]$ with label $c\in\Sigma$ is denoted by $q[i].\Delta[c]=q[j]$.
SA is the smallest DAG that satisfies:
(i) every node in $\mathcal{V}$ is reachable from $q[0]$;
(ii) outgoing edges of the same node have distinct character labels;
(iii) the concatenation of labels along any path from $q[0]$ to an accepting state $v\in F$ forms a suffix of the string; and
(iv) every suffix of the string can be represented by such a path.
For example, Fig.~\ref{fig:sa-single} shows the SA of string \kw{abbb}, where $F=\{q_4,q_5,q_6\}$.
The label sequence on the paths from $q_0$ to $q_4$ and $q_6$, and the two paths from $q_0$ to $q_5$, represent all suffixes.

\begin{algorithm}[t]
\small
\SetVline
\SetFuncSty{textsf}
\SetArgSty{textsf}
\caption{\kw{BuildSA {\color{gray}/ BuildMultiSA}}($\mathcal{T}$)}
\label{algo:sam-con-unified}
\Input{$\mathcal{T}=\{T_1,\ldots,T_n\}$ {\color{gray}and corresponding vectors $\{v_1,\ldots,v_n\}$}}
\Output{an SA {\color{gray}/ \kw{MultiSA}} $\mathcal{G}$}

\State{Create $q[0]$ with $q[0].len\leftarrow 0$; $q[0].link\leftarrow\emptyset$; {\color{gray} $q[0].\mathcal{O}\leftarrow\emptyset$}}

\For{each $T_j\in\mathcal{T}$}{
    \State{$last\leftarrow q[0]$}
    \For{each character $t_i$ of $T_j$ in order}{
        \State{Create a state $cur$ with $cur.len\leftarrow last.len+1$}
        \State{$p\leftarrow last$}
        \While{$p\neq\emptyset$ and $p.\Delta[t_i]=\emptyset$}{
            \State{$p.\Delta[t_i]\leftarrow cur$; $p\leftarrow p.link$}
        }

        {\color{gray}
        \State{$last.\mathcal{O}\leftarrow last.\mathcal{O}\cup\{v_j\}$}
        \If{$p=last$}{
            \State{$p'\leftarrow last.\Delta[t_i]$}

            \State{{\color{gray}\textbf{if} $last.len+1=p'.len$ \textbf{then} Remove $cur$; $last\leftarrow p'; \textbf{continue}$}}
            \State{Create a clone $p''$ of $p'$ with $p''.len\leftarrow last.len+1$}
            \State{$p'.link\leftarrow p''$}
            \While{$p\neq\emptyset$ and $p.\Delta[t_i]=p'$}{
                \State{$p.\Delta[t_i]\leftarrow p''$; $p\leftarrow p.link$}
            }
            \State{$last\leftarrow p''$; Remove $cur$; \textbf{continue}}

        }}
        \State{\textbf{if} $p=\emptyset$ \textbf{then} {{$cur.link\leftarrow 0$; $last\leftarrow cur$; \textbf{continue}}}}
            \State{$p'\leftarrow p.\Delta[t_i]$; $last\leftarrow cur$}
            \State{\textbf{if} $p.len+1=p'.len$ \textbf{then} {{$cur.link\leftarrow p'$}}}
            \Else{
                \State{Create a clone $p''$ of $p'$ with $p''.len\leftarrow p.len+1$}
                \State{$cur.link\leftarrow p''$; $p'.link\leftarrow p''$}
                \While{$p\neq\emptyset$ and $p.\Delta[t_i]=p'$}{
                    \State{$p.\Delta[t_i]\leftarrow p''$; $p\leftarrow p.link$}
                }
            }

    }
    {\color{gray}
    \State{$last.\mathcal{O}\leftarrow last.\mathcal{O}\cup\{v_j\}$}
    }
}

\State{$\mathcal{G}\leftarrow$ the set of all created, non-removed states}
{\color{gray}
\For{each state $p\in\mathcal{G}$ in descending order of $p.len$}{
    \State{\textbf{if} $p.link\neq\emptyset$ \textbf{then}
           $p.link.\mathcal{O}\leftarrow p.link.\mathcal{O}\cup p.\mathcal{O}$}
}
}

\State{\Return{$\mathcal{G}$}}
\end{algorithm}

For a string $T$, its suffix automaton (SA) $\mathcal{G}=(\mathcal{V},\mathcal{E})$ contains at most $2|T|-1$ states and $3|T|-4$ transitions, and can be constructed in $O(|T|)$ time~\cite{BlumerBHECS85}.
During construction, each state $p\in\mathcal{V}$ maintains its maximum represented string length $p.len$ and a suffix link $p.link$.
The latter points to the state representing the longest proper-suffix class of the strings represented by $p$; the initial state $q[0]$ has $q[0].link=\emptyset$.
The black statements in Algorithm~\ref{algo:sam-con-unified} outline single-string SA construction, i.e., here $n=1$, i.e., $\mathcal{T}=\{T\}$. 
Let $last$ be the state corresponding to the current prefix before inserting $t_i$; initially, $last=q[0]$ (line 3). 
For each character $t_i$ (line 4), we first create a new node $cur$ with $cur.len=last.len+1$ (line 5), and then trace suffix links from $last$ (line 6). 
As long as the current node has no transition labeled $t_i$, we add a transition to $cur$ and continue following suffix links (lines 7-8). 
If no such node $p$ exists, we set $cur.link\leftarrow q[0]$ and finish this iteration (line 18). Otherwise, let $p'=p.\Delta[t_i]$ (line 20). If $p.len+1=p'.len$, we set $cur.link\leftarrow p'$ (line 21). 
If not, we clone $p'$ into a new node $p''$, copy all information from $p'$ except setting $p''.len\leftarrow p.len+1$ (line 23), and then set both $cur.link$ and $p'.link$ to $p''$ (line 24). 
Finally, we continue tracing suffix links from $p$ and redirect every transition labeled $t_i$ that points to $p'$ to point to $p''$ instead (lines 25-26).
For example, the constructed SA of string \kw{abbb} is shown in Fig.~\ref{fig:sa-single}, where the represented substring set of each state is illustrated, e.g., $q_5$ represents \kw{abbb} and \kw{bbb}, corresponding to the two paths from $q_0$ to $q_5$.

\section{Limitations of Existing Approaches}
\label{sec:existing}

\begin{figure*}[t]
\begin{minipage}{0.22\linewidth}
\includegraphics[width=0.90\linewidth]{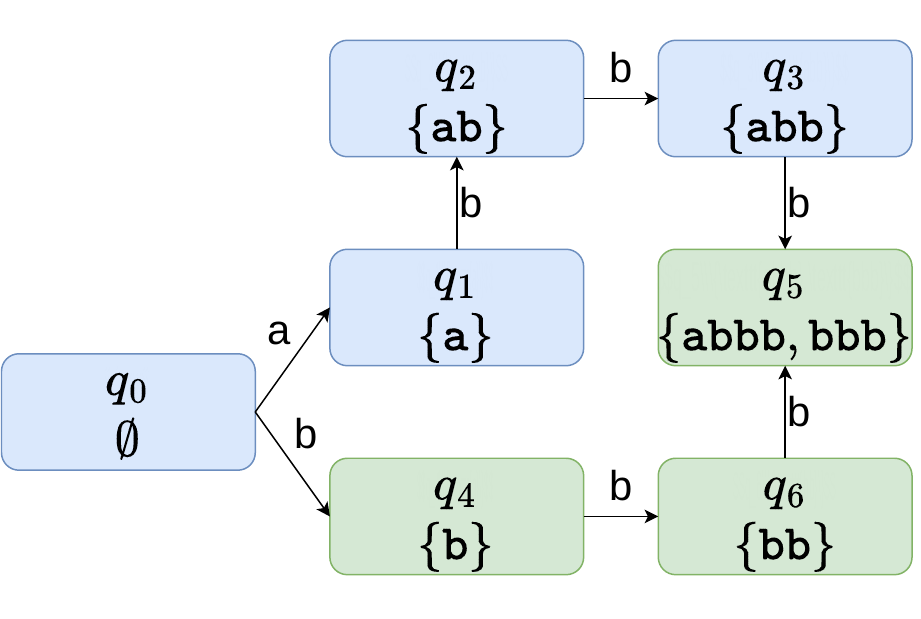}
\vspace{-4mm}
\caption{A SA of String \kw{abbb}}
\label{fig:sa-single}
\vspace{-5mm}
\end{minipage}
\hspace{2mm}
\begin{minipage}{0.30\linewidth}
\includegraphics[width=\linewidth]{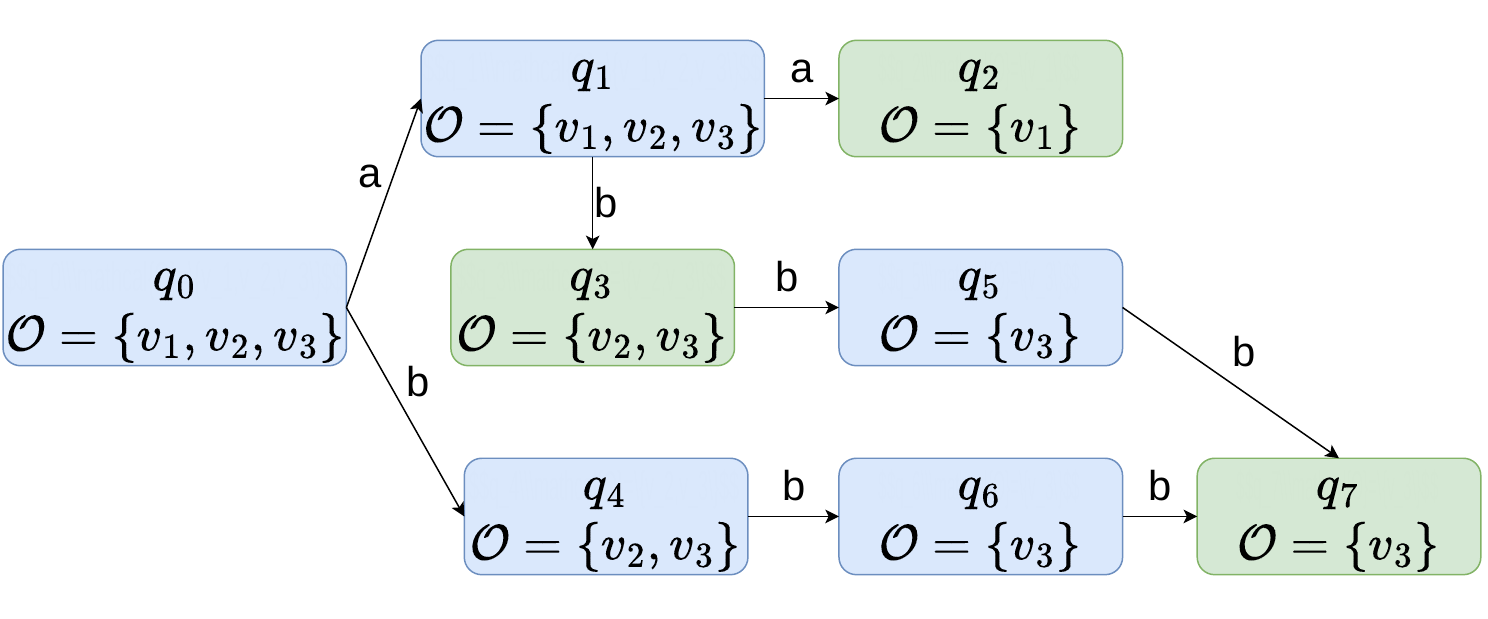}
\vspace{-4mm}
\caption{An ESA of Strings $\{s_1=\kw{aa},s_2=\kw{ab},s_3=\kw{abbb}\}$}
\label{fig:example-esa}
\vspace{-5mm}
\end{minipage}
\hspace{2mm}
\begin{minipage}{0.25\linewidth}
\includegraphics[width=\linewidth]{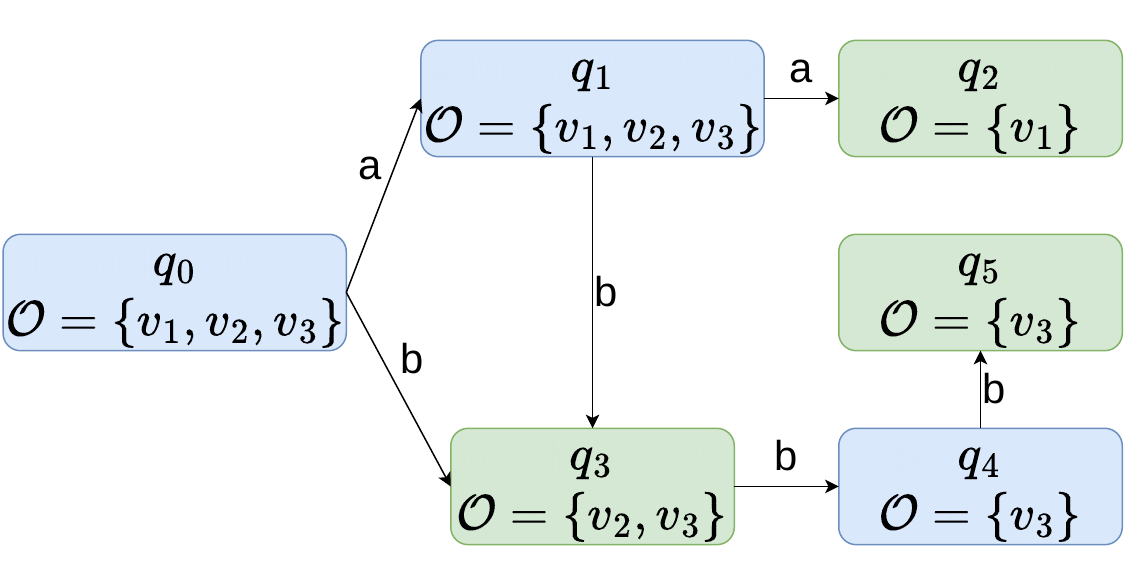}
\vspace{-4mm}
\caption{An \kw{MultiSA} of Strings $\{s_1=\kw{aa},s_2=\kw{ab},s_3=\kw{abbb}\}$}
\label{fig:example-multisa}
\vspace{-5mm}
\end{minipage}
\hspace{2mm}
\begin{minipage}{0.17\linewidth}
\includegraphics[width=0.98\linewidth]{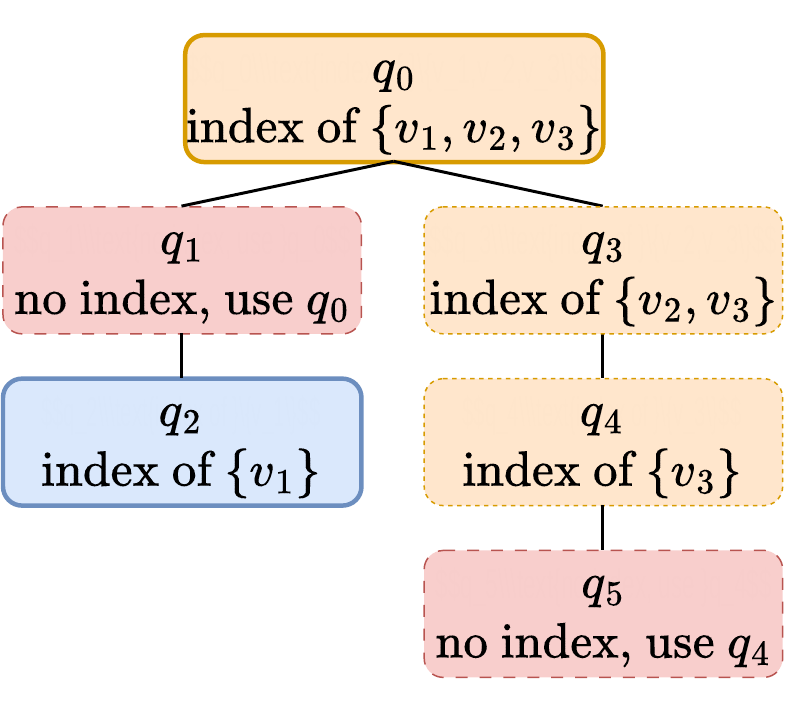}
\vspace{-4mm}
\caption{A Link Tree of \kw{MultiSA} in Fig.~\ref{fig:example-multisa}}
\label{fig:example-linktree}
\vspace{-5mm}
\end{minipage}
\end{figure*}

In this section, we first discuss \VectorMaton~\cite{abs-2603-01525}, the most closely related and state-of-the-art index for substring filtered ANN search, and then summarize why other filtered ANN approaches do not fully address our setting.

\stitle{\VectorMaton~\cite{abs-2603-01525}:}
\VectorMaton extends SA, which represents all substrings of a single string, to a collection of strings. Specifically, it constructs an enhanced suffix automaton (ESA), where each state represents a set of equivalent substrings having identical occurrence-position lists, with each occurrence recorded as a pair of sequence ID and ending position. This extension allows \VectorMaton to distinguish substrings that occur in the same set of sequences but at different positions, thereby maintaining the exact occurrence information across multiple strings. Similarly, the resulting ESA forms a directed acyclic graph (DAG) through character transitions. To support substring filtered $k$-ANN queries, \VectorMaton considers constructing an index containing the corresponding vectors of each state in ESA.
For index reuse, it exploits the descendant relationships in this transition DAG. Specifically, let $\mathcal{O}_x$ be the set of objects contained in state $x$ in ESA, for each state $x$, it selects the descendant $y$ with the largest indexed vector set as its inherited index, i.e., $\mathcal{O}_y\subseteq \mathcal{O}_x$, and constructs a local index only over the remaining vectors, i.e., vectors in $\mathcal{O}_x\setminus \mathcal{O}_y$. It further adopts a threshold-based skip-build strategy, which stores small residual vector sets directly rather than constructing HNSW graphs.

\VectorMaton has several issues, as illustrated below.

\noindent\underline{\ding{192} Automaton Issue:}
\VectorMaton groups substrings according to their complete occurrence-position lists. In our problem, however, the needed information is only the set of vector IDs whose strings contain the query string $s_q$; occurrence positions and multiplicities are not considered. Two states can therefore be equivalent for all possible future query outputs even when their occurrence-position lists differ.
For example, consider three objects $o_1=(v_1,\texttt{aa})$, $o_2=(v_2,\texttt{ab})$, and $o_3=(v_3,\texttt{abbb})$, the corresponding ESA is shown in Fig.~\ref{fig:example-esa}. Considering the substrings $\texttt{ab}$ (state $q_3$) and $\texttt{b}$ (state $q_4$), they have different occurrence-position lists, i.e., $\{(2,2),(3,2)\}$ and $\{(2,2),(3,2),(3,3),(3,4)\}$, respectively, and are therefore kept as two ESA states.
However, as $\mathcal{O}_{q_3}=\mathcal{O}_{q_4}$, for every continuation $x$, they induce the same vector set, i.e., $\mathcal{O}_{\texttt{ab}x}=\mathcal{O}_{\texttt{b}x}$ always holds.
The same holds for $\texttt{abb}$ (state $q_5$) and $\texttt{bb}$ (state $q_6$).
Hence, as shown in Fig.~\ref{fig:example-multisa}, the eight-state ESA in this example can be quotiented into six states by merging
$\{q_3,q_4\}$ and $\{q_5,q_6\}$,
without changing the answer to any substring-filtered query.
Therefore, an automaton built from position-list equivalence is correct, but its states are not minimized. Thus, the equivalent states in ESA add index cost, as \VectorMaton considers building an index for each state.

\noindent\underline{\ding{193} Index Reuse Issue:}
For each state $x$, \VectorMaton greedily selects an indexed descendant $y$ with the largest indexed vector set as its inherited index, and constructs a local index only over the residual vector set. Therefore, a query on $x$ requires two independent searches, one on its local index and the other on the inherited index of $y$, followed by merging their results. This introduces additional search and result-merging overhead compared with searching a single index over the entire vector set.

Furthermore, its index construction plan is heuristic and provides no global optimality guarantee. Specifically, \VectorMaton makes a greedy, state-wise reuse decision based on the size of a descendant's indexed set. An ancestor can reuse a descendant as a single index only when the descendant's complete candidate set is covered by that index, since  \VectorMaton limits query searches on at most two indices.
For example, consider ESA in Fig.~\ref{fig:example-esa}, $q_7$ first builds an index over $\{v_3\}$. States $q_5$ and $q_6$ can directly reuse this index. States $q_3$ and $q_4$ inherit the index of $q_7$ and each builds a residual index over $\{v_2\}$. State $q_2$ builds an index over $\{v_1\}$. For state $q_1$, the indexed sets of $q_7$ and $q_2$ have the same size, so either choice leaves a residual set of size two. However, $q_1$ is then represented by two indexes and cannot be reused by $q_0$ as a single complete index. Thus, $q_0$ must inherit a single-index descendant such as $q_7$ or $q_2$, leaving a residual set of size at least two.
Under this greedy plan, the index objects associated with $q_7$, $q_3$, $q_4$, $q_2$, $q_1$, and $q_0$ store 8 vectors in total. In contrast, if we build one complete index for $q_1$ over $\{v_1,v_2,v_3\}$, then $q_0$ can directly reuse the index of $q_1$. The resulting plan keeps five indices and stores 7 vectors in total.
As a result, locally attractive reuse decisions may lead to a suboptimal total index cost, leaving opportunities for further index sharing unexplored and potentially incurring unnecessary graph construction.

\stitle{String-Aware Retrieval and Issues:}
Dense-sparse ANN methods~\cite{ChenZCXZWHYLWLY24,abs-2410-20381,abs-2511-00855} jointly score vector similarity and lexical relevance. Although lexical representations can capture substrings, they provide only a soft matching signal rather than the exact Boolean condition that $s_q$ must occur as a contiguous substring. Thus, dense-sparse search cannot simultaneously guarantee both accurate filtering and high-quality $k$-ANN retrieval in our setting, since filtering needs higher weight on sparse vectors while $k$-ANN needs higher weight on dense vectors. Moreover, adapting these methods requires retrieving an enlarged candidate set and verifying the substring predicate afterward, which incurs cost on irrelevant candidates' distance computations and additional predicate verifications.

\stitle{General Filtered ANN Index and Issues:}
General filtered-ANN techniques include pre-filtering~\cite{WangYGJXLWGLXYY21,ZhangXCSXCCH00Y23,WeiWWLZ0C20}, post-filtering~\cite{WangYGJXLWGLXYY21,WeiWWLZ0C20,ZhanSWPLWCLPZC19}, and graph-search variants that incorporate the predicate during traversal~\cite{ZhangXCSXCCH00Y23,PatelKGZ24}. 
Although these methods usually have lower index cost, this advantage is often less decisive in practice than search efficiency.
By contrast, their search efficiency is highly sensitive to predicate selectivity. Pre-filtering becomes expensive when the qualifying set is large, whereas post-filtering and graph-based methods may visit many objects that do not satisfy the substring predicate. Since vector-distance evaluation dominates proximity graph search~\cite{GaoL23,YangLJZWSJW25}, such visits translate directly into redundant computations and unstable performance across query regimes.

\section{A New Multi-Proximity Graph Index}
\label{sec:overview}

To address the issues of string-aware retrievals and general filtered ANN indices for achieving optimal search performance on substring filtered ANN, an ideal scenario involves an index that, for each query substring $s_q$, can provide a separate proximity graph built only over vectors whose attribute contains $s_q$. Such a graph would avoid visiting non-qualifying nodes altogether. However, the number of distinct substrings across the dataset is quadratic in string length, i.e., $O(\sum_{i=1}^n|s_i|^2)$ for set $\{s_1,\cdots,s_n\}$, so building a graph per substring is infeasible in both space and build time.

The key observation, inspired by the suffix automaton, is that different substrings may induce exactly the same candidate object set. Therefore, instead of building one index for every distinct substring, it is sufficient to build graphs only for equivalence classes of substrings that share the same satisfying object set. However, the existing ESA does not guarantee to achieve a minimized state automaton; hence, to address \ding{192} Automaton Issue, we first design a state-minimized suffix automaton for our problem, called \kw{MultiSA}.

We discuss the \kw{MultiSA} with examples in detail. 

\stitle{\kw{MultiSA}:}
In \kw{MultiSA}, each state represents all substrings that produce the same object set after every possible continuation. 
Specifically, Let $F(p)=\{i\in[1,n]\mid p\text{ is a substring of }s_i\}.$
Two substrings $p$ and $q$ can share an automaton state only if $\forall x\in\Sigma^*,\;F(px)=F(qx).$ This condition includes the current output ($x=\emptyset$) and every possible continuation. Note that merely having $F(p)=F(q)$ is insufficient, since the two substrings may admit different future extensions.
For instance, the state $q_3$ and $q_4$ in ESA shown in Fig.~\ref{fig:example-esa} is merged as $q_3$ in \kw{MultiSA} shown in Fig.~\ref{fig:example-multisa}, since for every continuation $x\in\Sigma^*$, $\mathcal{O}_{\texttt{ab}x}=\mathcal{O}_{\texttt{b}x}$ always holds.
As in a suffix automaton, a substring is represented by a path from the initial state $q_0$, with edge labels spelling the substring.
The object set associated with a state $u$, denoted by $u.\mathcal{O}$, is the set of objects whose attributes contain $p$  as a substring.
For example, Fig.~\ref{fig:example-multisa} shows a \kw{MultiSA} constructed over the object set $\{(v_1,\texttt{aa}),(v_2,\texttt{ab}),(v_3,\texttt{abbb})\}$.
Node $q_3$ represents the substrings \kw{ab} and \kw{b}, corresponding to the two paths from $q_0$ to $q_3$. It represents the object set $\{v_2,v_3\}$, because both objects contain at least one substring represented by $q_3$.

The construction details, minimization procedure, and proof of \kw{MultiSA} are given in Section~\ref{sec:new-sam}.

Based on \kw{MultiSA}, we can design a multiple proximity graph index, \Surf, where each state in \kw{MultiSA}, built from the string set $\{s_i|o_i=(v_i,s_i)\in \mathcal{O}\}$, contains a proximity graph.
Specifically, to construct \Surf, we construct proximity graphs, e.g., HNSW~\cite{hnsw}, for $p.\mathcal{O}$ of every node $p$ in \kw{MultiSA}.

\stitle{Search Algorithm on \Surf:}
Given a query $q=(v_q,s_q)$, recall that in \kw{MultiSA}, each node represents the substrings spelled by the concatenation of edge labels along any path from $q_0$ to that node. Thus, to process the query, we start at $q_0$ and follow the edge labels to locate the node representing $s_q$, and then perform $k$-ANN search on the proximity graph associated with that node.
As shown in Algorithm~\ref{algo:search}, the search starts at $q_0$ in \kw{MultiSA} (line 1). We navigate $cur$ to the node representing $s_q$ through the transitions labeled $c_i$ of $s_q$ sequentially (lines 2, 4), and terminate the search immediately if such a node is not found (line 3).
Upon locating the node, we conduct a beam search on $G$ of this node to obtain the final $k$-ANN result (lines 5-6). Note that $G$ here is built exactly based on the set of objects that satisfy the query predicate, achieving the ideal search scenario we discussed before.

\begin{algorithm}[t]
\small  
    \SetVline 
    \SetFuncSty{textsf}
    \SetArgSty{textsf}
 \caption{\kw{SURFSearch}($\mathcal{I},v_q,s_q,k,w$)}
 \label{algo:search}
 \Input{index $\mathcal{I}$, query $q=(v_q,s_q)$, $k$ for top-$k$ and beam width $w$}
 \Output{$k$-ANN result}
 \State{$cur\leftarrow q[0]$}
 \For{each $c_i\in s_q$ in string order}
 {
    \State{\textbf{if} $cur.\Delta[c_i]=\emptyset$ \textbf{then} \Return{$\emptyset$}}
    \State{\textbf{else} $cur\leftarrow cur.\Delta[c_i]$}
 }
 \State{$G\leftarrow$ the proximity graph of $cur$ containing $cur.\mathcal{O}$}
 \State{\Return{\kw{Beam Search}($G,v_q,k,w$)}}
\end{algorithm}

However, although \kw{MultiSA} has minimized the number of states in ESA for lower index construction cost, both space consumption and construction time of \Surf are still up to a factor of the average string length compared with the original single proximity graph.
Therefore, to make the index scalable in practice, we consider compressing it.
\VectorMaton provides a way to compress the proximity graphs among states in an automaton; however, as discussed in \ding{193} index reuse issue, it does not provide any theoretical guarantee on compressed quality and does not fully utilize the relationships among states as shown below.

The key observation is that, for any two substrings $s_1$ and $s_2$, if $s_1$ is a suffix of $s_2$, then the object set corresponding to $s_2$ must be a subset of that corresponding to $s_1$. The reason is that any object whose string contains $s_2$ as a substring must also contain $s_1$, since $s_1$ is itself a substring of $s_2$. Based on this property, as we will prove in Section~\ref{sec:compressed} later, the nodes in \kw{MultiSA} induce a tree structure, which we refer to as \emph{link tree}. We discuss it in detail with examples.

\stitle{Link Tree:}
A link tree is a tree rooted at $q[0]$ and can be represented as a graph $\mathbb{G}=(\mathbb{V},\mathbb{E})$, where $\mathbb{V}=\mathcal{V}$ is \kw{MultiSA} state set. An edge from $p$ to $u$ in $\mathbb{E}$ is a parent-child edge, indicating the strings represented by $p$ are the longest suffixes of those represented by $u$. Thus, $u.\mathcal{O}\subseteq p.\mathcal{O}$ always holds.
For example, the link tree corresponding to \kw{MultiSA} shown in Fig.~\ref{fig:example-multisa} is illustrated in Fig.~\ref{fig:example-linktree}. Consider nodes $q_3$, $q_4$, and $q_5$, which represent strings $\{\kw{b},\kw{ab}\}$, $\{\kw{bb,abb}\}$, and $\{\kw{bbb,abbb}\}$, respectively. Since both $\{\kw{b},\kw{ab}\}$ and $\{\kw{bb,abb}\}$ are suffixes of $\{\kw{bbb,abbb}\}$, but $\{\kw{bb,abb}\}$ have the larger length, the parent of $q_5$ is $q_4$.

Based on the link tree, in which the parent of each node corresponds to a superset of the child's object set, we introduce two complementary compression rules.
(1) Path compression stores a sequence of logical graph versions in one physical graph by incrementally inserting the set differences between adjacent retained states.
(2) Ancestor compression avoids materializing a logical index for a state when its candidate set is sufficiently large relative to the candidate set of an already retained ancestor. The serving ancestor is not necessarily the direct parent; it can be the nearest retained ancestor on the link-tree path.

We describe these two rules in detail below.

\stitle{Path Compression:}
Consider a path $p_1,p_2,\cdots,p_m$ from a lower-depth node to a higher-depth node in the link tree. Since $p_m.\mathcal{O}\subseteq p_{m-1}.\mathcal{O}\subseteq \cdots \subseteq p_1.\mathcal{O},$
the proximity graphs on these nodes can be constructed incrementally from $p_m$ to $p_1$, i.e., for each node $p_i$ with $i<m$, we derive it from the graph of $p_{i+1}$ by inserting the vectors in $p_i.\mathcal{O}\setminus p_{i+1}.\mathcal{O}$.
In this way, we do not need to build $m$ separate proximity graphs for $p_1,p_2,\cdots,p_m$. Instead, we only need to record the changes during this incremental construction process, i.e., the edges added and the edges pruned, so that the proximity graph of each node can be efficiently recovered from these records.

\begin{figure}
    \centering
    \includegraphics[width=0.9\linewidth]{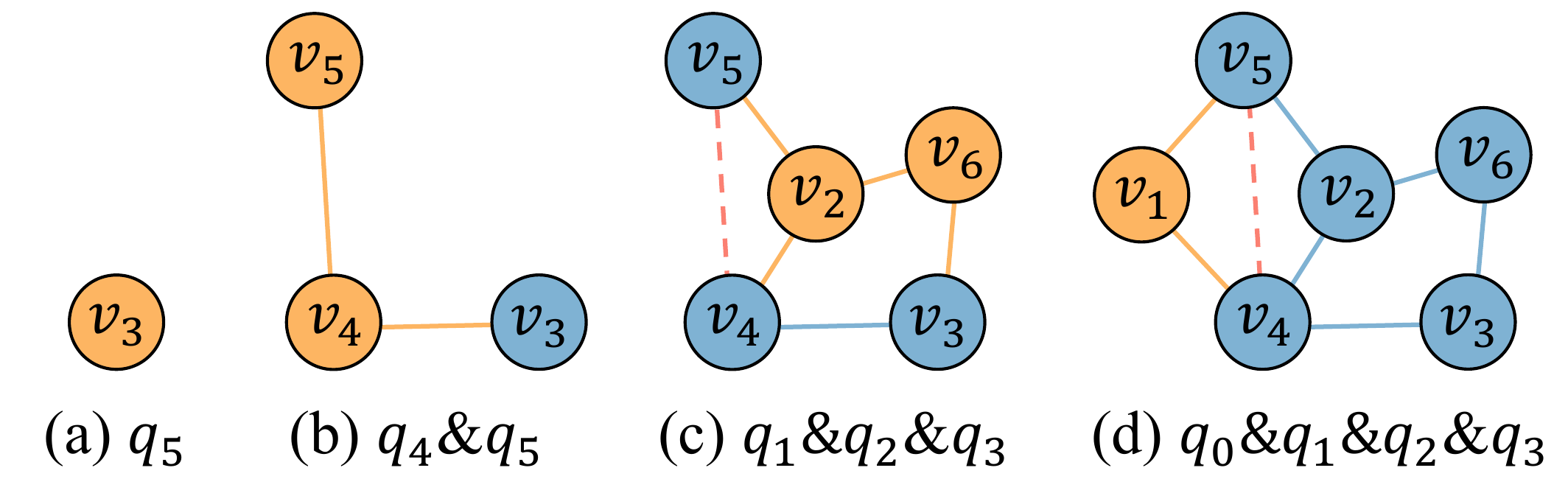}
    \vspace{-3mm}
    \caption{An Example of Path Compression}
    \label{fig:example-path}
    \vspace{-3mm}
\end{figure}

\begin{example}
Adding objects $o_4=\{v_4,\kw{abb}\}$, $o_5=\{v_5,\kw{abb}\}$ and $o_6=\{v_6,\kw{ab}\}$ into \kw{MultiSA} shown in Fig.~\ref{fig:example-multisa}, the only differences are the object set of states. Considering four states $q_0,q_3,q_4,q_5$, we have  $q_0.\mathcal{O}=\{v_1,v_2,\cdots,v_6\}$, $q_3.\mathcal{O}=\{v_2,v_3,\cdots,v_6\}$, $q_4.\mathcal{O}=\{v_3,v_4,v_5\}$, and $q_5.\mathcal{O}=\{v_3\}$.
As illustrated in Fig.~\ref{fig:example-path}, we first construct the index for $q_5$ with vector $v_3$. We then insert $v_4$ and $v_5$ to obtain the index for $q_4$. Next, for $q_3$, we insert $v_2$ and $v_6$, during which the newly added edges prune edge $(v_4,v_5)$. Finally, we insert $v_1$ to obtain the index for $q_0$. Throughout this process, we record all index updates, and pruned edges are not physically removed. As a result, we only need to store the recorded index evolution along the path $q_5,q_4,q_3,q_0$, from which the graph of any node on the path can be extracted.
\end{example}

\stitle{Ancestor Compression:}
Let $u$ be a state and let $a$ be an ancestor of $u$ in the link tree whose logical index is retained. Since $u.\mathcal{O}\subseteq a.\mathcal{O}$, the index of $a$ can serve queries on $u$ by verifying the original substring predicate after retrieving. We skip index construction for $u$ when ${|u.\mathcal{O}|}/{|a.\mathcal{O}|}\ge\alpha$. Importantly, $a$ is not necessarily a direct parent of $u$. Here, ${|u.\mathcal{O}|}/{|a.\mathcal{O}|}$ is the selectivity, and $\alpha$ is predefined and is set as $0.3$ by default.

This compression rule needs to be integrated with methods from the post-filtering strategy. According to the results in existing work~\cite{XuGGLJ24,CaiSCZ24,ZuoQZLD24,ZhangXCSXCCH00Y23}, the post-filtering demonstrates effectiveness, even similar to search on a dedicated proximity graph index, when the selectivity of the query filter is high.
In particular, when a search is required on such state $u$, we employ beam search on the proximity graph of $a$ to retrieve all $w$ results in $S$, rather than just obtaining the top-$k$ results. Next, we check all $w$ vectors to ascertain if their string attributes satisfy the filter to derive the final top-$k$ results.
For example, in Fig.~\ref{fig:example-linktree}, we do not build an index for $q_1$, but instead reuse the index of $q_0$ to support queries on them, since the selectivity $|q_1.\mathcal{O}|/|q_0.\mathcal{O}|=1$ is high.
For search, we perform beam search with beam width $w>k$ on the index of $q_0$, and then discard nodes in $q_0.\mathcal{O}\setminus q_1.\mathcal{O}$ ($\emptyset$ here) if it appears among the $w$ retrieved results before returning the final top-$k$ answers.

To apply the two compression rules optimally for minimizing the index cost of \Surf, we propose a dynamic programming method in Section~\ref{sec:optimal-compress} that jointly considers both rules and computes the optimal compression plan.
Based on these rules, together with \Surf search algorithm and the post-filtering strategy used for parent node compression, we develop \kw{ComSURF}. \kw{ComSURF} significantly reduces index size and building time, while preserving, and often even improving, search performance. Details are given in Section~\ref{sec:compressed}.

\section{Optimal Compression Strategy}
\label{sec:optimal-compress}

In this section, we jointly optimize path compression and ancestor compression on the link tree of \kw{MultiSA}. We formulate the problem as an exact dynamic program (DP) and then exploit the monotonicity of candidate-set sizes to obtain an efficient implementation.

Let $w(u)=|u.\mathcal{O}|$ be the candidate-set size of an active state $u$, i.e., a state for which an index is considered. States with $w(u)\le  ef_{\min}$ are handled by exact scanning and are excluded from the graph-planning tree. Since candidate sets are nested along link-tree edges, $w(u)\le w(parent(u))$. The remaining active states form a rooted tree containing the root whenever graph indexing is enabled.
Each active state has one of three roles:
\ding{172} \textbf{Skipped state}: no logical index is retained at $u$; its query is served by the nearest retained ancestor.
\ding{173} \textbf{Path anchor}: $u$ starts a physical proximity-graph path and contributes $w(u)$ to the index-cost proxy.
\ding{174} \textbf{Path continuation}: $u$ retains a logical graph version by extending the physical path of its direct parent.
The distinction between a logical index and a physical graph is important. A skipped state is served by the nearest retained ancestor $a$, which may be a path anchor or an internal logical version. Its eligibility condition is $w(u)\ge \alpha w(a),$ where $\alpha\in[0,1]$ is the selectivity threshold. At query time, the system searches the version associated with this actual owner and applies the original substring predicate to the retrieved candidates. A path continuation, in contrast, must be connected to its direct parent because the physical graph is built by incremental insertion along adjacent retained states.

Our optimization objective is the sum of candidate-set sizes of all physical path anchors, which captures the number of vector insertions across independently materialized paths.

We then consider the transition equations between these states.
For a node $u$, let $I[u]$ denote the minimum additional cost in the subtree of $u$ when $u$ retains a logical graph version and continues the physical path of its parent. Let $A[u]=w(u)+I[u]$ be the cost when $u$ starts a new physical path. Finally, let $F[u,a]$ be the minimum cost of the subtree rooted at $u$ when $u$ is skipped and is served by its nearest retained ancestor $a$. This state is feasible only if $w(u)\ge\alpha w(a)$.
\ding{192} If $u$ is skipped, each child can either start a new path or remain skipped under the same serving ancestor: $F[u,a]=\sum_{v\in child(u)}\min\left\{A[v],F[v,a]\right\},$
where the second term $F[v,a]$ is available only when $w(v)\ge\alpha w(a)$.
\ding{193} If $u$ is retained, define $\mu(u,v)=\min\left\{A[v],F[v,u]\right\},$ where $F[v,u]$ is considered only if $w(v)\ge\alpha w(u)$, and let $S_u=\sum_{v\in child(u)}\mu(u,v).$
At most one direct child can continue the same physical path. Choosing such a child $v$ replaces $\mu(u,v)$ by $I[v]$. Hence, $I[u]=S_u-\max\left(0,\max_{v\in child(u)}\{\mu(u,v)-I[v]\}\right),$ and $A[u]=w(u)+I[u].$
For a leaf, $I[u]=0$ and $A[u]=w(u)$. The root cannot be served by an ancestor, so the optimal cost is $A[root]$.
This formulation is exact, but explicitly maintaining all feasible ancestor contexts can require quadratic space and time on a long tree. We therefore remove the ancestor identity from the DP state.

The serving ancestor affects a skipped subtree only through the threshold $\alpha w(a)$. Let $t$ denote this threshold, and define $H_u(t)$ as the minimum cost of the subtree rooted at $u$ when $u$ is not connected to the path above it and may either start a new path or be skipped under an external retained ancestor with threshold $t$. Then $H_u(t)=\begin{cases}\min\left\{A[u],\displaystyle\sum_{v\in child(u)}H_v(t)\right\},& t\le w(u),\\[2mm]A[u],&t>w(u).\end{cases}$.
In particular, $\mu(u,v)=H_v(\alpha w(u))$. This recurrence is equivalent to the ancestor-context DP, but represents all possible serving ancestors through one scalar threshold.
Each $H_u$ is a nondecreasing step function, and its breakpoints occur only at candidate-set sizes of states in the subtree. We represent $H_u$ using a finite event set $E_u$, where each event $(k,\Delta)\in E_u$ records a positive increment $\Delta$ that takes effect when the threshold $t$ becomes strictly larger than $k$, i.e., $H_u(t)=\sum_{(k,\Delta)\in E_u:\,k<t}\Delta.$
We store a consumed prefix as a base value and keep only future events in an ordered map.

\begin{algorithm}[t]
\small  
    \SetVline 
    \SetFuncSty{textsf}
    \SetArgSty{textsf}
 \caption{\kw{CompressRule}($\mathbb{G},\alpha$)}
 \label{algo:dp}
    \Input{a link tree $\mathbb{G}$, and threshold $\alpha$}
\Output{the optimal compression plan}
\For{each node $u$ in bottom-up order}{
    \If{$child(u)=\emptyset$}{
        \State{$I[u]\leftarrow 0$; $A[u]\leftarrow w(u)$; $E_u\leftarrow\{(w(u),w(u))\}$}
        \State{\textbf{continue}}
    }
    \For{each child $v$ of $u$}{
        \State{Consume from $E_v$ all events with $k<\alpha w(u)$ into $base[v]$}
        \State{$\mu(u,v)\leftarrow base[v]$}
    }
    \State{$S_u\leftarrow\sum_{v\in child(u)}\mu(u,v)$}
    \State{$pick[u]\leftarrow\arg\max_{v\in child(u)}\{\mu(u,v)-I[v]\}$ if this maximum is positive; otherwise $pick[u]\leftarrow\emptyset$}
    \State{$I[u]\leftarrow S_u-\max(0,\max_{v\in child(u)}\{\mu(u,v)-I[v]\})$}
    \State{$A[u]\leftarrow w(u)+I[u]$}
    \State{Merge the remaining child event maps using small-to-large}
    \State{Cap the event sum at $A[u]-S_u$, and add the residual at key $w(u)$ if needed}
    \State{$b[u]\leftarrow$ largest key in the resulting event map}
}
\State{Reconstruct roles top-down, carrying the nearest retained owner $a$}
\State{Skip $u$ iff $\alpha w(a)\le b[u]$; otherwise start a new path}
\State{\Return{$pick$, $b$, and the reconstructed path and owner assignments}}
\end{algorithm}

As shown in Algorithm~\ref{algo:dp} with input of a link tree $\mathbb{G}$ and parameter $\alpha$, we process each state $u$ bottom-up (line 1), and initialize at leaf nodes (lines 2-4). 
We first query each child function at $t_u=\alpha w(u)$, then consume events with keys strictly smaller than $t_u$ into the child's base (lines 6-7). The sum of the resulting child bases is $S_u$ (line 8), and the largest saving determines the child selected for path continuation (line 9). The node then obtains $I[u]$ and $A[u]$ from the recurrence above (lines 10-11). The remaining child events are merged, capped so that their total does not exceed $A[u]-S_u$, and, if necessary, completed by an event at key $w(u)$ (lines 12-13).

For every active node with positive candidate count, the final function has a positive jump, i.e., $A[u]>S_u$. To see this, the path saving at $u$ is strictly smaller than $w(u)$; otherwise, a child of equal weight would have to attain its anchor cost at the same threshold, contradicting the inductive strict inequality for that child. Thus, the capped event map is nonempty, and its largest key $b[u]$ is well-defined.
Let $b[u]$ be the largest key of the final event map of $u$. Since the event representation is nondecreasing and its final value is $A[u]$, we have $H_u(t)<A[u]\Leftrightarrow t\le b[u].$
The left-hand side means that reusing the given ancestor is strictly cheaper than starting a new path. During top-down reconstruction, we therefore carry the nearest retained owner $a$. A node is skipped exactly when $\alpha w(a)\le b[u]$; otherwise it starts a new path and becomes the serving owner for its skipped descendants. If a retained node is selected as the continuation child of its parent, it is attached to that parent's physical path. 

\begin{theorem}
Algorithm~\ref{algo:dp} returns a feasible plan minimizing the total candidate-set size of all physical path anchors.
\end{theorem}

\proofsketch
The ancestor-context recurrence enumerates every feasible role of a node and, when the node is retained, every possible direct child that continues its physical path. Once this choice is fixed, the remaining child subtrees are independent. The threshold-function recurrence is an exact reparameterization of the ancestor context because the latter is used only through $\alpha w(a)$. The bottom-up event operations evaluate this recurrence exactly, while the boundary $b[u]$ preserves precisely the strict comparison needed for top-down reconstruction. Induction from leaves to the root establishes optimality. 
\eop

\stitle{Time Complexity:}
Let $m$ be the number of active nodes. Each node contributes at most one new event. Events consumed by an ancestor or removed by the cap never reappear. The event maps are merged small-to-large, selecting the heavy child by active-subtree size. Thus, each surviving event is migrated across at most $O(\log m)$ light edges, and each map operation costs $O(\log m)$. The total planning complexity is therefore $O(m\log^2m)$ time.

\section{Compressed Index of \Surf}
\label{sec:compressed}

In this section, we detail the construction of \kw{ComSURF}. We first present the smallest deterministic state-output automaton for our problem and its construction details. We then introduce the link tree, which exposes the containment relationships used by the two compression rules. Finally, we explain how path compression and ancestor compression are realized by the physical index and how the optimal plan is reconstructed.

\subsection{State-Minimized Suffix Automaton}
\label{sec:new-sam}

The multi-string suffix automaton in \cite{abs-2603-01525} groups substrings by their occurrence-position lists. For substring filtered ANN, however, a query needs only the set of object IDs whose strings contain the substring; occurrence positions are not part of the consideration. We therefore minimize states with respect to the output function $F(p)$ defined in Section~\ref{sec:overview}, while preserving all future transitions.

For two substrings $p$ and $q$, define $p\equiv_F q$ iff $\forall x\in\Sigma^*,\;F(px)=F(qx).$
This is the coarsest equivalence relation that permits deterministic state merging. Equality of the current candidate sets alone is not sufficient; for example, for a dataset containing the single string \kw{ab}, both \kw{a} and \kw{b} output the same object ID, but appending \kw{b} distinguishes them because \kw{ab} exists whereas \kw{bb} does not.
Let $\delta(u,c)$ be the transition from state $u$ on character $c$, and let $[v]$ denote the equivalence class
of state $v$ in the minimized automaton.
Since character transitions form a DAG, we compute the
quotient bottom-up in reverse topological order.
The signature of a state is defined as $sig(u)=\left(u.\mathcal{O},\left\{(c,[\delta(u,c)])\mid c\in\Sigma,\ \delta(u,c)\neq\emptyset\right\}\right).$
Two states are merged exactly when their complete signatures match. We use hashing to efficiently locate candidate signatures and verify the full ID arrays and transition lists before merging.

\begin{theorem}
The quotient obtained by merging states with identical signatures is the minimal deterministic state-output automaton.
\label{theo:output-minimal}
\end{theorem}

\proofsketch
If two substrings reach the same state in any correct deterministic state-output automaton, then every common continuation must produce the same output. Hence, no machine can merge substrings that are not equivalent under $\equiv_F$. Conversely, $\equiv_F$ is a right congruence: if $p\equiv_F q$, then $pc\equiv_F qc$ for every character $c$. Therefore, the equivalence classes define a deterministic quotient whose output is well-defined and equals $F(p)$. The quotient reaches the lower bound and is minimal.
\eop

\sloppy
We construct \kw{MultiSA} by inserting the object strings $\mathcal{T}=\{s_1,\ldots,s_n\}$ one by one with gray lines in Algorithm~\ref{algo:sam-con-unified}. Before inserting a new string, we reset $last$ to the initial state $q_0$ (line 3), thereby preventing transitions from representing substrings that cross string boundaries. During character insertion, if $last$ already has a transition labeled by the new character (line 12), we reuse its target when its maximum length is $last.len+1$; otherwise, we split the target state as in the standard SA construction (lines 13-17).
Meanwhile, the vector associated with each string is attached to the end states of its prefixes, and a reverse traversal of suffix links propagates these vectors to all states representing their substrings (lines 28-29). 

\stitle{Time Complexity:}
Let $L=\sum_{i=1}^{n}|s_i|$ be the total length of all input strings, $L_{\max}=\max_i|s_i|$, and $\Sigma$ be the  alphabet.
The \kw{MultiSA} construction creates $Q=O(L)$ states and $T=O(L|\Sigma|)$ transitions, and takes $O(L)$ time under constant-time transition operations and amortized suffix-link traversal.
Let $B=\sum_{u\in\mathcal{V}}|u.\mathcal{O}|$ denote the total number of candidate-ID entries across all base states.
Given sorted candidate-ID arrays, computing state signatures and verifying candidate matches takes expected $O(B+T+Q)$ time using hashing. To process states in reverse topological order, we bucket-sort them by their maximum represented substring length, which takes $O(Q+L_{\max})$ time.
Sorting outgoing transitions by character takes $O(T+Q)$ time for a bounded alphabet.
Therefore, the total expected time complexity is $O(B+T+Q+L_{\max})$.
Combining base construction and minimization, the overall expected preprocessing time is $O(B+L|\Sigma|)$.
Since $B=O(L^{1.5})$ in the worst case, the overall expected time is $O(L^{1.5})$ for a bounded alphabet.

\subsection{Tree Structure among \kw{MultiSA} Nodes}
\label{sec:link-tree}

After state minimization, multiple original states may be merged into a single state in \kw{MultiSA}. Therefore, the parent of a minimized state cannot be directly defined as the suffix link of an arbitrary original state in its equivalence class. We define the parent by skipping the original suffix links within the same equivalence class.

Specifically, let $\mathcal{G}_0$ denote the automaton before minimization, and let $\pi:\mathcal{V}_0\rightarrow\mathcal{V}$ map each original state to its corresponding state in \kw{MultiSA}.
For each minimized state $p\neq q_0$, let $u_p$ be the original state in $\pi^{-1}(p)$ with the smallest maximum represented substring length.
We define its parent as $p.link=\pi(u_p.link_0),$ where $link_0$ denotes the original suffix-link function before minimization.
That is, the parent of $p$ is the minimized state reached by following the original suffix link from the shortest state in its equivalence class.
The root state $q_0$ has no parent.

The following theorem establishes that these parent relations form a tree and preserve the candidate-set containment property required by both compression rules.

\begin{theorem}
The parent relations defined above form a tree rooted at $q_0$.
Moreover, if $q$ is an ancestor of $p$ in this tree, then $p.\mathcal{O}\subseteq q.\mathcal{O}$.
\label{coro:multisa}
\end{theorem}

\proofsketch
We first establish that every equivalence class under $\equiv_F$ forms a contiguous chain under the original suffix links.
For an object string $s_i$ and a substring $x$, define $R_i(x)=\{z\in\Sigma^*\mid xz\text{ is a substring of }s_i\}.$
By the definition of $\equiv_F$, if $p\equiv_F q$, then $R_i(p)=R_i(q)$ for every object $i$.
Let $e_i(x)$ denote the earliest ending position of an occurrence of $x$ in $s_i$.
The maximum length of a valid  continuation in $R_i(x)$ is $|s_i|-e_i(x)$. Therefore, $e_i(p)=e_i(q).$ Consequently, if $|p|<|q|$, then $p$ is a suffix of $q$.
Now consider any suffix $r$ of $q$ such that $|p|<|r|<|q|$. Since $p$ is a suffix of $r$, and $r$ is a suffix of $q$, for every object $i$ we have $R_i(q)\subseteq R_i(r)\subseteq R_i(p).$
As $R_i(p)=R_i(q)$, it follows that $R_i(r)=R_i(p)$ for every object $i$. Hence, $p\equiv_F r\equiv_F q$.
In a suffix automaton, the states representing successive suffixes of a string are encountered along its suffix-link chain. Thus, all original states belonging to the same equivalence class form a contiguous chain under the original suffix links.
Furthermore, the state with the smallest maximum represented substring length in each equivalence class is unique. Otherwise, two distinct states with the same minimum length would represent the same shortest substring, which cannot reach two distinct states in a deterministic automaton.
We can therefore contract each equivalence class into one state in \kw{MultiSA}.
By the construction of the original multi-string suffix automaton, every non-root state has exactly one suffix link, and following this link strictly decreases its maximum represented substring length.
Hence, repeatedly following suffix links from any non-root state eventually reaches $q_0$, and the original suffix-link relations form a rooted tree.
Since each equivalence class forms a contiguous chain in this tree, contracting these chains preserves the rooted-tree structure.
Moreover, the parent of each contracted state is precisely the first state outside its equivalence class encountered along the original suffix-link chain, which agrees with the definition of $p.link$ above.
Therefore, the parent relations form a rooted tree with $q_0$ as its root.

It remains to establish candidate-set containment.
For any original suffix-link edge $u\rightarrow u.link_0$, every substring represented by $u.link_0$ is a suffix of a substring represented by $u$. Therefore, every object containing a substring represented by $u$ also contains its represented suffix, yielding $u.\mathcal{O}\subseteq u.link_0.\mathcal{O}.$
Applying this property along the original suffix-link chain, and then contracting the chains, gives $p.\mathcal{O}\subseteq q.\mathcal{O}$,
whenever $q$ is an ancestor of $p$ in the link tree.
\eop

We refer to graph $\mathbb{G}$ as the \textbf{link tree} of a \kw{MultiSA} $\mathcal{G}$.
It is obtained by contracting the original suffix-link tree according to the equivalence classes produced by state minimization, and therefore requires no separate tree construction procedure.

\subsection{Index Construction}
\label{sec:comsurf-index}

We now describe how to construct \kw{ComSURF} using the optimal compression plan of Section~\ref{sec:optimal-compress}. Path compression represents the logical graphs of adjacent retained states along a link-tree path in one physical graph, while ancestor compression skips selected states and serves their queries using the nearest retained ancestor.

\stitle{Path Compression:}
Consider a path $p_1,p_2,\ldots,p_m$ from ancestor to descendant, where $p_i.link=p_{i-1}$ for $i\ge 2$. By Theorem~\ref{coro:multisa},
$p_m.\mathcal{O}\subseteq\cdots\subseteq p_1.\mathcal{O}$.
Rather than building $m$ independent proximity graphs, we first construct $G_m$ on $p_m.\mathcal{O}$, and then obtain $G_i$ from $G_{i+1}$ for $i=m-1,\ldots,1$ by inserting the vectors in $p_i.\mathcal{O}\setminus p_{i+1}.\mathcal{O}$.
Each newly created edge is labeled by the version $p_i$ in which it becomes visible. If graph pruning removes an older edge from the logical graph of $p_i$, we retain the edge physically and attach a negative label $\neg p_i$. Hence, the final physical graph stored at $p_1$ encodes every logical graph $G_i$ on the path; a query for $p_i$ simply traverses the edges visible in version $p_i$.

\begin{algorithm}[t]
\small  
    \SetVline 
    \SetFuncSty{textsf}
    \SetArgSty{textsf}
 \caption{\kw{PathConstruct}($G_i,p_i,p_{j},efc,M$)}
 \label{algo:path-pg}
 \Input{a proximity graph $G_i$ of node $p_i$ and node $p_{j}$}
 \Output{a proximity graph $G_{j}$ of node $p_{j}$}
 \State{$G_{j}\leftarrow G_i$}
 \For{each vector $v\in p_{j}.\mathcal{O}\setminus p_i.\mathcal{O}$}
 {
    \State{$C\leftarrow$\kw{BeamSearch}($G_{j},v,efc,efc$)}
    \State{utilize RNG pruning strategy to make $|C|\le M$}
    \For{each node $u$ in $C$}
    {
        \State{add edges $(u,v)$ and $(v,u)$ with labeled $p_j$ to $G_{j}$}
        \If{$|N_{G_{j}}(u)| > M$}
        {
            \State{$w\leftarrow $ node pruned by RNG pruning strategy in $N_{G_{j}}(u)$}
            \State{add a label $\neg p_j$ to edge $(u,w)$ in $G_{j}$}
        }
    }
 }
 \State{\Return{$G_{j}$}}
\end{algorithm}

To add vectors to a proximity graph for constructing a new one, navigable small world graphs, e.g., HNSW~\cite{hnsw}, a state-of-the-art approach for $k$-ANN queries, naturally support vector insertion. Here, we illustrate the process using HNSW as an example to present the details of the construction of such a proximity graph $G_j$ for $p_j$ based on $G_{i}$ of $p_i$, as depicted in Algorithm~\ref{algo:path-pg}, where $efc$ and $M$ are two hyperparameters in HNSW construction for controlling the size of search beam width and node degree in index respectively.
Let $G_j$ initially equal $G_i$ (line 1), followed by the insertion of each vector in $p_{j}.\mathcal{O}\setminus p_i.\mathcal{O}$ (line 2). When inserting a vector $v$, we first obtain its $k$-ANN as the candidate neighbor set (line 3) and then prune them to ensure the size does not exceed $M$ using RNG pruning strategy (line 4). This strategy involves evaluating nodes $u\in C$ of node $v$, pruning $u$ if there exists $w \in C$ such that $\delta(v, w) < \delta(v, u)$ and $\delta(u, w) < \delta(v, u)$ simultaneously hold ~\cite{nsg,hnsw}. If $|C|>M$ holds after pruning, the node $u\in C$ with the largest distance to $v$ will be removed until $C\le M$.
Next, for each node $u\in C$ (line 5), we add edges from $u$ to $v$ and from $v$ to $u$ (line 6). The addition of these edges may cause the degree of $u$ to surpass $M$. Thus, we also apply RNG pruning strategy to limit the size of its neighbor list (line 8). The removed edge will be marked $\neg p_j$, which is not truly removed from $G_j$, but excluded from beam searches of node $p_j$ and line 3.

It is worth noting that incorporating other state-of-the-art methods into Algorithm~\ref{algo:path-pg}, such as NSG~\cite{nsg}, $\tau$-MNG~\cite{taumg}, and ALMG~\cite{ALMG} for path compression, is feasible. The key distinction lies in these graphs not inherently facilitating vector insertions. Nonetheless, various methods are available to aid in proximity graph maintenance~\cite{abs-2105-09613,WangYGJXLWGLXYY21,abs-2503-00402,XieYL25}. Therefore, by leveraging these techniques, any proximity graph can be integrated seamlessly.

\begin{theorem}
For a path $p_1,p_2,\cdots,p_m$ with $p_i.\mathcal{O}\subseteq p_{i-1}.\mathcal{O}$ for all $i\ge 2$, the versioned graphs constructed from $p_m$ to $p_1$ using Algorithm~\ref{algo:path-pg} has $O(M\cdot |p_1.\mathcal{O}|)$ physical adjacency records.
\label{theo:compress-path}
\end{theorem}

\proofsketch
The first graph $G_m$ stores $O(M|p_m.\mathcal{O}|)$ edges. Each subsequently inserted vector adds at most $O(M)$ new adjacency records; pruning only marks previously created records rather than creating additional graph edges. Since the inserted sets along the path are disjoint and their union is $p_1.\mathcal{O}\setminus p_m.\mathcal{O}$, the total number of records is $O(M|p_m.\mathcal{O}|)+O(M|p_1.\mathcal{O}\setminus p_m.\mathcal{O}|)=O(M|p_1.\mathcal{O}|)$.
\eop

Thus, path compression reduces index-size from $O\!\left(M\sum_{i=1}^m|p_i.\mathcal{O}|\right)$ for separate graphs to $O(M|p_1.\mathcal{O}|)$ for a shared, versioned graph.

\stitle{Ancestor Compression:}
Let $u$ be a state and $a$ its nearest retained ancestor in the link tree. Since $u.\mathcal{O}\subseteq a.\mathcal{O}$, queries on $u$ can use the logical graph version of $a$, followed by verification of the original substring predicate. We may skip the index at $u$ when $   \frac{|u.\mathcal{O}|}{|a.\mathcal{O}|}\ge\alpha.$

We next give the construction details of \kw{ComSURF}.

\stitle{Index Construction of \kw{ComSURF}:}
In summary, to build the compressed \Surf, \kw{ComSURF}, we first construct \kw{MultiSA} using Algorithm~\ref{algo:sam-con-unified} and then minimize it via hashing, as discussed earlier. Next, we derive the link tree by suffix links of each state in \kw{MultiSA}. 
For states with $|p.\mathcal{O}|\le ef_{\min}$, we retain the exact candidate sets for direct scanning and omit proximity graph construction. These states are excluded from graph planning without altering their outputs in \kw{MultiSA}. We then run Algorithm~\ref{algo:dp} on remaining active link tree to obtain the retained states, compressed paths, and serving ancestors. Finally, we construct one versioned graph for each selected path.

\kw{ComSURF} can support insertions, deletions, and string-attribute updates as discussed below.

\stitle{Dynamic Maintenance:}
We separate maintenance of the logical automaton from that of the physical proximity graphs. For an insertion $(v,s)$ with a fresh ID, we incrementally update the underlying multi-string automaton using Algorithm~\ref{algo:sam-con-unified}. The updated automaton remains exact, while its state-minimization and graph-compression plans are reconciled during periodic consolidation. A new string can split existing states, so the updated automaton may no longer share the compression plan of the physical index. We therefore maintain a stable graph-index snapshot and make recent insertions searchable through a delta layer until consolidation. Queries search the snapshot and the delta layer, verify the substring predicate where necessary, and merge their results. Delta states with at most $ef_{\min}$ candidates use exact scanning; larger states may maintain their own proximity graphs. 
Deletions are initially recorded in tombstone bitmaps. Tombstoned vectors are excluded from returned results but may temporarily remain as intermediate graph-traversal connectors. A string-attribute update is processed as a tombstone for the old version followed by insertion of a new version with a fresh internal ID. 
Periodically, we merge the delta layer into the graph-index snapshot, derive the current link tree, and recompute the compression plan using Algorithm~\ref{algo:dp}. For insertion-only batches, the incrementally maintained \kw{MultiSA} can be used directly. When deletions or replacements occur, we rebuild \kw{MultiSA} if multiple states have empty ID sets.

\section{Search Algorithms on \kw{ComSURF}}
\label{sec:comsurf-search}

In this section, we describe query processing on \kw{ComSURF}. We first show how the compression plan obtained in Section~\ref{sec:optimal-compress} determines the logical graph used by a substring filtered ANN query. We then extend the search to SQL \kw{LIKE} predicates.

\stitle{Interpreting the Compression Plan:}
Algorithm~\ref{algo:dp} records $pick[u]$, the child selected to continue the physical path of a retained state $u$, and $b[u]$, the largest threshold for which skipping $u$ is strictly cheaper than starting a new path. Starting from the retained root $q[0]$, we traverse the active link tree top-down while carrying the nearest retained ancestor $a$. For a child $u$ of a retained node $p$, if $u=pick[p]$, we retain $u$ as the next logical version on $p$'s physical path. Otherwise, whether $p$ is retained or skipped, we skip $u$ precisely when $\alpha w(a)\le b[u]$; if this inequality fails, $u$ starts a new physical path. Here $a=p$ when $p$ is retained, and otherwise it is the nearest retained ancestor inherited from above.
For each active state $u$, the resulting layout identifies its \emph{serving owner} $owner(u)$: $owner(u)=u$ if $u$ is retained, and otherwise $owner(u)$ is its nearest retained ancestor. For a retained state $a$, let $anchor(a)$ be the first (lowest-depth) state of its compressed path, which stores the corresponding physical graph. Both mappings are derived from $pick$ and $b$ during the top-down reconstruction in Section~\ref{sec:optimal-compress}.

\stitle{Substring Filtered Queries:}
For a query $q=(v_q,s_q)$, we start at $q[0]$ and follow the character transitions of $s_q$ in \kw{MultiSA}. If any transition is missing, the qualifying set is empty and we return no result. Otherwise, we reach a state $u$ with $u.\mathcal{O}=\mathcal{O}_{s_q}$. There are three possibilities.
\ding{192} For $|u.\mathcal{O}|\le ef_{\min}$, we directly scan its small candidate set, which gives the exact result.
\ding{193} If $u$ is retained, we run beam search on its logical graph version. Every traversed vector then satisfies the substring predicate. \ding{194} If $u$ is skipped, we search the logical version of $a=owner(u)$ and verify the retrieved candidates against $u.\mathcal{O}$ (equivalently, against the original substring predicate). For this last case, we retain up to $w$ candidates when employing beam search before filtering, and return the closest $k$ qualifying candidates found.

Locating the matching state takes $O(|s_q|)$ transition steps. Subsequent work is either an exact scan of at most $ef_{\min}$ vectors or one beam search over the selected logical graph version, plus at most $w$ predicate checks for a skipped state. No additional graph materialization is required during search.

\stitle{\kw{LIKE} Predicate Queries:}
A SQL \kw{LIKE} predicate $l_q$ may contain \kw{\%}, matching zero or more characters, and \kw{\_}, matching exactly one character. A substring filter $s_q$ is the special case ``\kw{LIKE} \%$s_q$\%''. We first handle the substring-matching interpretation of a wildcard pattern and then enforce any prefix or suffix anchoring required by the complete SQL predicate.

Remove the leading and trailing \kw{\%} symbols, if present, to obtain a core pattern $l_q$.
As shown in Algorithm~\ref{algo:like-search}, we maintain a set $C$ of possible \kw{MultiSA} states, initially $\{q[0]\}$ (line 1). On an ordinary character $c$, every state follows its $c$-labeled transition (line 5). On \kw{\_}, we follow all one-character outgoing transitions (line 3). On \kw{\%}, we include every state reachable by zero or more character transitions, including the current states (line 4). Thus, $C$ contains all possible automaton endpoints for substrings matching the processed core pattern. If $C$ becomes empty, no object can satisfy the predicate.

Many reachable states are redundant. If states $p,u\in C$ satisfy that $p$ is an ancestor of $u$ in the link tree, then $u.\mathcal{O}\subseteq p.\mathcal{O}$ by Theorem~\ref{coro:multisa}. More importantly, this dominance persists after any common continuation, as formalized below. Therefore, after each wildcard-navigation step, we remove every state having a distinct ancestor also in $C$ (line 7), reducing the number of states carried forward without losing any qualifying object.

\begin{algorithm}[t]
\small  
    \SetVline 
    \SetFuncSty{textsf}
    \SetArgSty{textsf}
 \caption{\kw{LIKESearch}($\mathcal{I},v_q,l_q,k,w$)}
 \label{algo:like-search}
 \Input{index $\mathcal{I}$, query $q=(v_q,l_q)$, $k$ for top-$k$ and beam width $w$}
 \Output{$k$-ANN result}
 \State{$C\leftarrow \{q[0]\}$}
 \For{each $c_i\in l_q$ in string order}
 {
 \State{\textbf{if} $c_i=\texttt{\_}$ \textbf{then} $C\leftarrow \bigcup_{\sigma\in \Sigma}\{u|p\in C\wedge p.\Delta[\sigma]=u\}$}
    \State{\textbf{else if} $c_i=\texttt{\%}$ \textbf{then} $C\leftarrow \{u|p\in C\wedge u$ can be reached by $p$$\}$}
    \State{\textbf{else} $C\leftarrow \{u|p\in C\wedge p.\Delta[c_i]=u\}$}
    \State{\textbf{if} $C=\emptyset$ \textbf{then} \Return{$\emptyset$}}
    \State{$C\leftarrow \{u\in C\mid \not\exists p\in C,p\neq u\wedge p\text{ is ancestor of }u\text{ in link tree}\}$}
 }
 \State{$R\leftarrow \emptyset$}
 \For{each $p\in C$}
 {
    \State{$G\leftarrow$ the proximity graph of $p$ containing $p.\mathcal{O}$}
    \State{$R\leftarrow R\cup${\kw{Beam Search}($G,v_q,k,w$)}}
 }
 \State{\Return{top $k$ vectors with smallest distance to $v_q$ in $R$}}
\end{algorithm}

\begin{theorem}
\label{theo:like-dominance}
Let $p$ be an ancestor of $u$ in the link tree of \kw{MultiSA}. If the transition $u.\Delta[c]$ exists for a character $c$, then $p.\Delta[c]$ also exists and is an ancestor of, or equal to, $u.\Delta[c]$ in the link tree. Hence, removing $u$ when $p$ is also in the current state set preserves the union of qualifying object sets after any subsequent sequence of ordinary characters and the wildcards \kw{\%} and \kw{\_}.
\end{theorem}

\proofsketch
A substring represented by $p$ is a suffix of a substring represented by $u$. Appending same character $c$ preserves this suffix relationship. Thus, if a transition on $c$ exists from $u$, the corresponding transition also exists from $p$, and their target states preserve the ancestor relation in the link tree. Applying this argument repeatedly gives the claim for any fixed continuation; \kw{\_} and \kw{\%} operations take unions over one-character and arbitrary-length continuations, respectively, so the containment remains valid under either wildcard. Since $u.\mathcal{O}\subseteq p.\mathcal{O}$, discarding $u$ does not change the union represented by the current states or by any future update.
\eop

After the core pattern has been processed, the union $\bigcup_{u\in C}u.\mathcal{O}$ contains every object satisfying the full \kw{LIKE} predicate. If the pattern begins and ends with \kw{\%}, this union is exactly the qualifying object set. Otherwise, it can be a superset, because \kw{MultiSA} recognizes substring occurrences rather than enforcing the beginning or end of an object string. In either case, Algorithm~\ref{algo:like-search} runs beam search on each remaining state with a beam width of $w$, deduplicates the results, verifies the complete \kw{LIKE} predicate against each object's original string, and returns the closest $k$ qualifying candidates.

\begin{example}
Consider the three strings $s_1=\kw{aa}$, $s_2=\kw{ab}$, and $s_3=\kw{abbb}$ in Fig.~\ref{fig:example-multisa}. For the predicate ``\kw{LIKE} \kw{\%a\%b\_\%}'', we first follow the transition on \kw{a}, and $C=\{q_1\}$. 
Next, expanding all states reachable from $q_1$ through zero or more transitions for the internal $\kw{\%}$ yields $C=\{q_1,q_2,q_3,q_4,q_5\}$.
$q_2$ will be removed because $q_1$ is its ancestor, while $q_4$ and $q_5$ are removed because $q_3$ is an ancestor of both. Thus, $C=\{q_1,q_3\}$.
Following the transition on $\kw{b}$ gives $C=\{q_3,q_4\}$, where $q_4$ is subsequently removed because $q_3$ is its ancestor.
Finally, expanding one outgoing transition for \kw{\_}, $C=\{q_4\}$, and the union of the resulting states' candidate sets is $\{v_3\}$.
\end{example}

\begin{table}[t]
  \centering
  \caption{Statistics of Datasets}
  \vspace{-4mm}
  \label{tab:dataset}
  {\small
  \setlength{\tabcolsep}{3pt}
  \resizebox{\linewidth}{!}{
  \begin{tabular}{|l|r|r|r|r|l|l|c|}
    \hline
    \textbf{Dataset} & \textbf{Dim.} & \textbf{\#vectors} & \textbf{\#queries} & \textbf{$\overline{|s_i|}$} & \textbf{String Src.} & \textbf{Vector Type} & \textbf{Source} \\
    \hline \hline
    \SIFT         & 128  & 1,000,000 & 10,000 & 19.99 & Synthetic & Image         & \cite{sift} \\ \hline
    \GIST         & 960  & 1,000,000 & 1,000  & 20.01 & Synthetic & Image         & \cite{sift} \\ \hline
    \Redcaps      & 512  & 1,000,000 & 1,000  & 29.99 & Synthetic & Image \& Text & \cite{DesaiKA021} \\ \hline
    \kw{Deep}   & 256  & 1,000,000 & 1,000  & 19.98 & Synthetic & Image         & \cite{wang2021comprehensive} \\ \hline
    \arXiv        & 768  & 157,606   & 1,000  & 99.08 & Real text & Text          & \cite{AspectPaperEmbeddings} \\ \hline
    \kw{Laion}  & 512  & 1,000,448 & 10,000 & 34.11 & Real text & Image & \cite{abs-2111-02114} \\ \hline
    \kw{DBLP}     & 768  & 1,000,000 & 10,000 & 38.99 & Real text & Text          & \cite{DBLPXML} \\ \hline
    \kw{IMDB}     & 1,024 & 1,000,000 & 10,000 & 29.45 & Real text & Text            & \cite{IMDBNonCommercialDatasets} \\ \hline
  \end{tabular}
  }}
  \vspace{-4mm}
\end{table}

\section{Experiments}
\label{sec:exp}

\begin{figure*}[!t]
  \centering
  \includegraphics[width=0.95\textwidth]{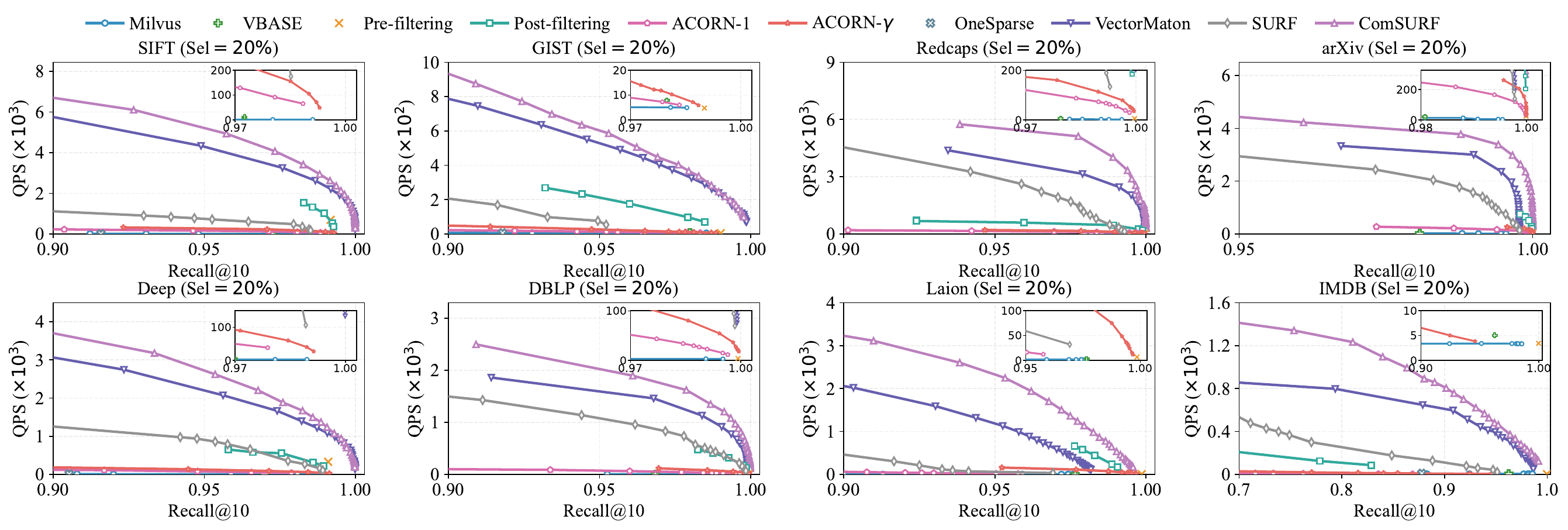}

  \includegraphics[width=0.95\textwidth]{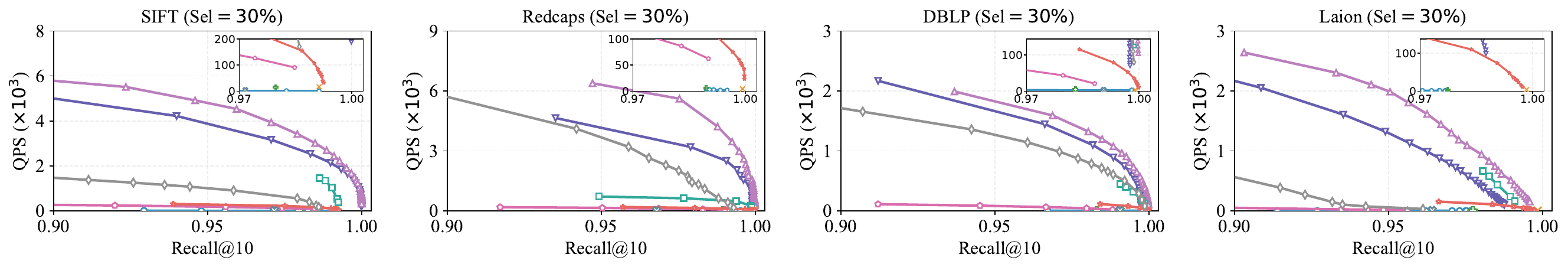}

  \includegraphics[width=0.475\textwidth]{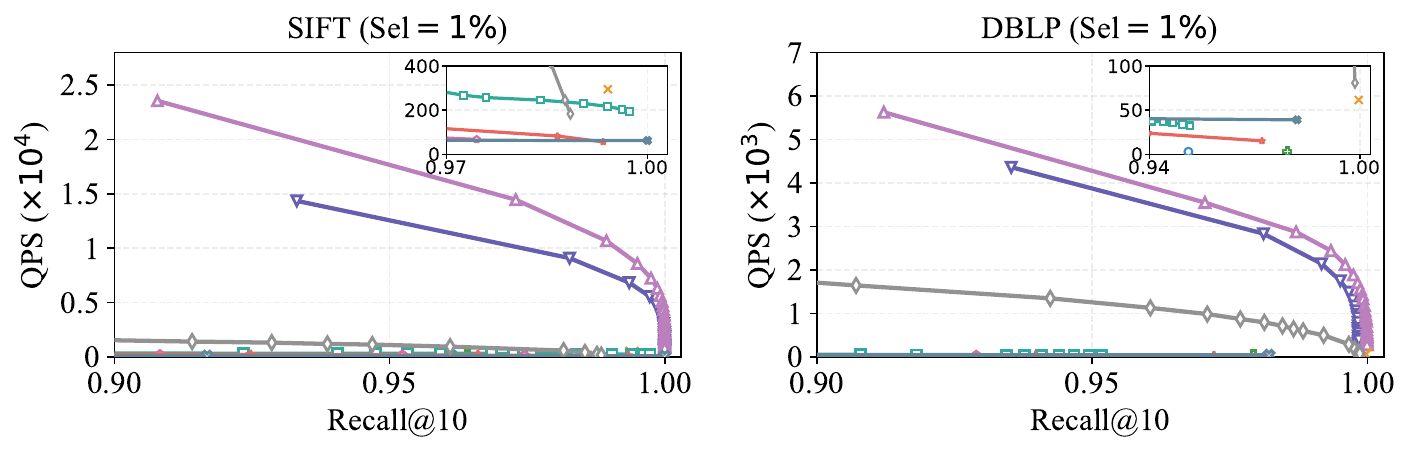}
  \hspace{-1mm}
  \includegraphics[width=0.475\textwidth]{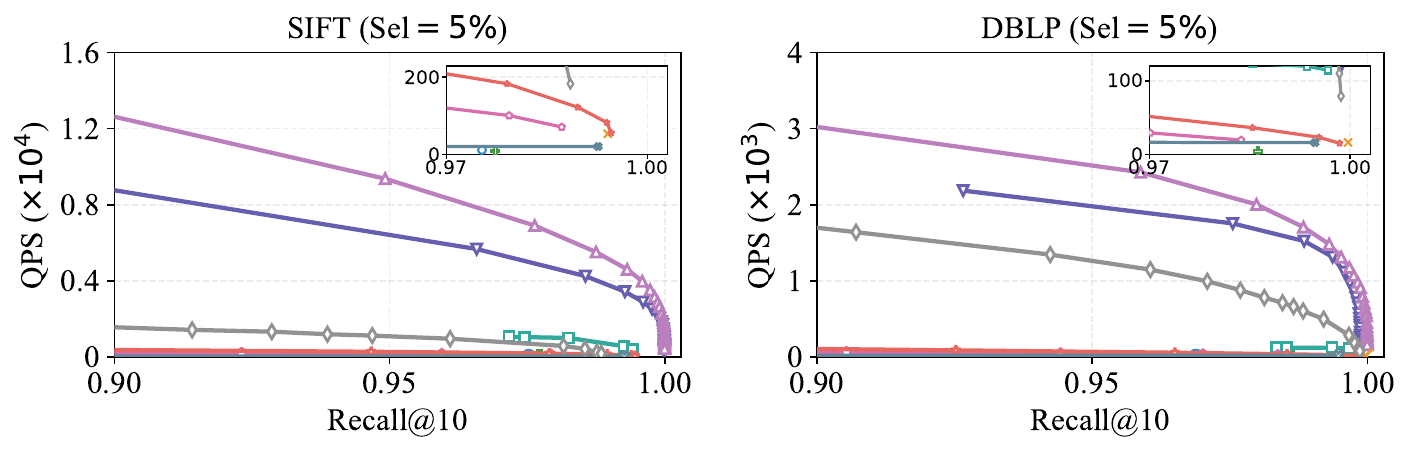}
  \vspace{-5mm}
  \caption{Overall search performance comparisons under different selectivities (Exp. 1)}
  \label{fig:qps-recall-0.2}
  \vspace{-4mm}
\end{figure*}

\stitle{Datasets:}
We employ 8 real-world public datasets along with their queries in our experiments.
The summary can be found in Table~\ref{tab:dataset}, including vector dimensionality (dim.), the number of base vectors (\#vectors), the number of query vectors (\#queries), the average length of string attributes in the base set ($\overline{|s_i|}$), the source of string attributes (string src.), the vector type, and the dataset source.

\noindent\underline{Base Data.}
For \arXiv and \kw{DBLP} datasets, vectors are generated using {SPECTER2}~\cite{Singh2022SciRepEvalAM}; for \kw{IMDB} dataset, vectors are generated using {e5-large-v2}~\cite{wang2022text}. The remaining datasets provide vectors directly.
In every dataset, each object $o_i$ is associated with a string attribute $s_i \in \Sigma^*$.
Datasets \arXiv, \kw{Laion1M}, \kw{DBLP}, and \kw{IMDB} contain native text metadata, from which we derive $s_i$ as follows: paper titles for \arXiv and \kw{DBLP}; a truncated caption with a fixed maximum length for \kw{Laion1M}; and, for \kw{IMDB}, the concatenation of the person name and the two occupations with the longest tenure.
For the remaining datasets, we synthetically generate $s_i$ by sampling characters from an alphabet $\Sigma$ consisting of lowercase letters, digits, and punctuation symbols.
Unless otherwise specified, we use a Zipf distribution for synthetic string generation.
Table~\ref{tab:dataset} reports $\overline{|s_i|}$ for each dataset, and the {String Src.} column indicates whether $s_i$ is \textit{synthetic} or derived from \textit{real-text} metadata.

\noindent\underline{Query Data.}
Unless otherwise specified, we use the standard query set provided by each dataset, which is disjoint from the base vectors.
To generate a string predicate for a query vector $v_q$ while controlling selectivity, we first build a \kw{MultiSA} over all object strings $\{s_i\}$.
Each state $p$ is associated with a set $p.\mathcal{O}$ of objects whose string attributes contain a substring represented by $p$ (as defined in Section~\ref{sec:overview}).
Therefore, any representative substring $s_q$ from state $p$ has selectivity
$Sel(s_q)=|p.\mathcal{O}|/|\mathcal{O}|$.
For overall-performance evaluation, we sample \kw{MultiSA} states such that the resulting queries have an average selectivity around $20\%$ or $30\%$.
Given a selected state $p$, we instantiate a concrete predicate by first sampling a length
$\ell \in \{p.\textit{link}.\textit{len}+1, \ldots, p.\textit{len}\}$,
and then uniformly sampling one substring of length $\ell$ represented by $p$ as the filter string $s_q$.

\noindent\textbf{Algorithms:}
We compare our approaches, \Surf/\kw{ComSURF}, with the following seven baselines. For each algorithm, we use recommended parameters if provided, and tune the remaining parameters on each dataset to obtain a Pareto-optimal recall-QPS curve.

\textbf{(1)} Pre-filtering~\cite{WangYGJXLWGLXYY21}:
We first materialize the filtered candidate set $\mathcal{O}{s_q}$ by locating $s_q$ in \kw{MultiSA}. We then compute distances from $v_q$ to all candidates in $\mathcal{O}{s_q}$ and return the top-$k$ results.
\textbf{(2)} Post-filtering~\cite{WangYGJXLWGLXYY21}:
We build an HNSW index on the vectors with $M=32$ and $ef_c=256$. At query time, we retrieve $k'$ candidates ($k' > k$) using HNSW search and then apply the substring predicate, returning the first $k$ qualifying results.
\textbf{(3)} \Milvus~\cite{WangYGJXLWGLXYY21}:
We build an HNSW index with $M=32$ and $ef_c=256$. Since Milvus supports \texttt{LIKE} predicates in ANN search, for substring filtering we issue \texttt{text LIKE "\%$s_q$\%"}, where string attributes are stored as \texttt{text}; for general \texttt{LIKE} evaluation, we directly use the corresponding \texttt{LIKE} predicate in \Milvus.
\textbf{(4)} \ACORN-$1$ and \ACORN-$\gamma$~\cite{PatelKGZ24}:
\ACORN provides two variants. Following prior studies on filtered ANN search~\cite{XuGGLJ24,CaiSCZ24,XieYSL26}, we evaluate both using $M=32$, $M_{\beta}=64$, and $ef_c=256$. For \ACORN-$1$, we fix $\gamma=1$; for \ACORN-$\gamma$, we set $\gamma$ adaptively based on $Sel(s_q)$.
\textbf{(5)} \VBASE~\cite{ZhangXCSXCCH00Y23}:
We use \VBASE with an HNSW index on the vector column (constructed with $M=32$ and $ef_c=256$). Queries are issued in SQL by combining a substring predicate on \texttt{text} (via \texttt{LIKE}) with a $k$-NN ordering on \texttt{vec} (via \texttt{<->}).
\textbf{(6)} \kw{OneSparse}~\cite{ChenZCXZWHYLWLY24}:
We use \kw{OneSparse} to compute a hybrid score that combines dense and sparse similarities, retrieve the top-$k'$ candidates under this score, and then apply post filtering to return the final $k$ results. Following~\cite{ChenZCXZWHYLWLY24}, the hybrid score is
$\alpha\delta(v_q,v_i) + (1-\alpha)\delta(u_q,u_i)$,
where $v$ and $u$ denote the dense and sparse representations, respectively.
We tune the weight $\alpha$ and use $\alpha=0.05$, which yields the best performance in our experiments.
\textbf{(7)} \VectorMaton~\cite{abs-2603-01525}:
We build \VectorMaton using the enhanced suffix automaton (ESA) to organize pattern-equivalent vector subsets and construct state indexes with index reuse and skip-build strategies. For fair comparison, we use the same HNSW construction parameters as our approaches, i.e., $M=32$, $ef_c=256$, and $T=200$. 
At query time, we traverse the ESAM to locate the state corresponding to $s_q$, search its local index and inherited index (if any), and merge their results to obtain the final top-$k$ results.
For \kw{LIKE} predicate, as \VectorMaton only supports \kw{\_} wildcard, we only compare it with others under this predicate.
\textbf{(8)} \Surf/\kw{ComSURF} (ours):
For fair comparison, we use the same HNSW construction parameters as above (i.e., $M=32$ and $ef_c=256$) for all proximity graphs in SURF/ComSURF.
Since the fully uncompressed \Surf index is prohibitively large, we evaluate \Surf with $ef_{\min}=2,000$ for a reasonable memory occupation by default.
Unless otherwise specified, for \kw{ComSURF}, we set the parent node compression threshold to $\alpha=0.3$ and set $ef_{\min}=200$.

\noindent\textbf{Performance Metrics:}
Following prior studies on filtered ANN search~\cite{XuGGLJ24,CaiSCZ24,XieYSL26,ZuoQZLD24},
we measure efficiency using QPS (queries per second) and accuracy using recall.
QPS is the number of queries processed per second, and recall is defined in Section~\ref{sec:problem}.

\noindent\textbf{Experimental Environments:}
Experiments are conducted on a server equipped with an Intel(R) Xeon(R) Gold 6240 CPU @ 2.60GHz (72 threads) and 376\,GB of RAM, running Ubuntu 22.04.5 LTS.
All methods are implemented in C++17 and compiled with \texttt{-Ofast}.

For index construction, we report build time and index size on the same machine using 64 threads.
For search, we set $k=10$ by default and evaluate query performance using a single thread, following prior work~\cite{XieYSL26,XuGGLJ24,ZuoQZLD24,CaiSCZ24}.
To reach a target recall, we tune query-time parameters in each method: for post-filtering, we increase the candidate budget $k'$; for other methods, we increase the beam-search width $w$ (or the corresponding search parameter) until the desired recall is achieved.

\stitle{Exp. 1: Overall Query Performance under Different Selectivities.}
We evaluate the overall query performance of \kw{ComSURF} against all baselines under different selectivities.
Fig.~\ref{fig:qps-recall-0.2} reports the QPS-recall trade-offs on all eight datasets at $Sel(s_q)\approx 0.2$, and the results on \SIFT, \Redcaps, \kw{DBLP}, and \kw{Laion} datasets at $Sel(s_q)\approx 0.3$.
We further evaluate low-selectivity queries with $Sel(s_q)\approx 0.01$ and $0.05$ on \SIFT and \kw{DBLP} datasets.

At moderate selectivities, \kw{ComSURF} achieves higher QPS than the baselines at comparable recall.
Compared with the state-of-the-art \VectorMaton, at selectivity 0.2, \kw{ComSURF} achieves $1.16$--$1.99\times$ higher QPS at $recall@10 \ge 0.95$ across all eight datasets; at $recall@10\ge 0.99$, \kw{ComSURF} achieves $1.14$--$1.43\times$ higher QPS on \SIFT, \Redcaps, \arXiv, \kw{Deep}, and \kw{DBLP}, while the two methods perform comparably on \GIST.
At selectivity 0.3, \kw{ComSURF} achieves $1.10$--$1.76\times$ higher QPS at $recall@10 \ge 0.95$, and remains $1.28$--$1.38\times$ faster at $recall@10 \ge 0.99$ on the three datasets where both methods reach this recall level.
Moreover, \kw{ComSURF} achieves higher maximum recall on several datasets where \kw{VectorMaton} cannot reach 0.99, particularly \kw{Laion} and \kw{IMDB}.
Compared with the other baselines, at $recall@10\ge 0.99$ and selectivity 0.2, \kw{ComSURF} achieves $3.2$--$13.5\times$ higher QPS than Post-filtering, $2.4$--$9.9\times$ than \Surf, $20.5$--$53.1\times$ than \kw{ACORN}-$1$, and $11.0$--$45.5\times$ than \kw{ACORN}-$\gamma$.
The improvement over \Milvus is even larger.
Similar advantages are observed at selectivity 0.3.
The performance improvement of \kw{ComSURF} over \Surf comes from both index compression and adaptive query processing. \kw{ComSURF} combines pre-filtering for small candidate sets, post-filtering with retained ancestor graphs for skipped states, and dedicated graph search for other states.
The comparison is also affected by memory constraints. The fully uncompressed \Surf index causes out-of-memory failures on many datasets when $ef_{\min}=0$. We therefore set $ef_{\min}=2,000$ to make index construction feasible, requiring exact scans for states with at most 2,000 candidates. This constraint further degrades query performance by replacing graph search with potentially more expensive distance computations.

At low selectivities, \kw{ComSURF} continues to demonstrate strong performance.
Compared with \VectorMaton, \kw{ComSURF} achieves $1.25$--$1.59\times$ higher QPS at $recall@10 \ge 0.95$ under selectivity 0.01, and $1.22$--$1.38\times$ under selectivity 0.05.
At $recall@10\ge 0.99$, it achieves $1.13$--$1.34\times$ higher QPS than \VectorMaton across the two datasets and both selectivities.
Compared with the other baselines, at $recall@10\ge 0.95$, \kw{ComSURF} achieves $2.1$--$15.2\times$ higher QPS than \Surf and up to $100.5\times$ and $149.8\times$ higher QPS than Post-filtering and Pre-filtering, respectively.
Compared with moderate selectivities, Post-filtering degrades substantially because fewer traversed vertices satisfy the filter, resulting in more wasted exploration.
Pre-filtering becomes relatively more competitive as the candidate sets shrink, but remains considerably slower than \kw{ComSURF}.

\begin{table}[t]
  \centering
  \caption{Index construction time in seconds (Exp. 2)}
  \vspace{-4mm}
  \label{tab:build-time}
  {\small
  \setlength{\tabcolsep}{2pt}
  \resizebox{\linewidth}{!}{
  \begin{tabular}{|l|r|r|r|r|r|r|r|}
    \hline
        \textbf{Dataset} & \textbf{\Surf} & \textbf{\kw{ComSURF}} & \textbf{\VectorMaton} & \textbf{\ACORN-1} & \textbf{\ACORN-$\gamma$} & \kw{HNSW} & \textbf{\kw{OneSparse}} \\
        \hline \hline
            \SIFT        & 2,676  & 416 & 2,372  & 9   & 2,611  & 26  & 137 \\ \hline
      \GIST        & 8,031  & 2,647 & 9,819 & 40  & 7,735  & 134 & 70  \\ \hline
      \Redcaps     & 4,146  & 1,300 & 3,695 & 24  & 5,478  & 51  & 56 \\ \hline
            \kw{Deep}    & 3,895  & 964 &4,168 & 15  & 4,862  & 62  & 140 \\ \hline
      \arXiv       & 898    & 271 &682   & 4   & 1,194  & 7   & 21 \\ \hline
      \kw{Laion}   & 5,627  & 2,521 & 5,747 & 23  & 6,154  & 59  & 121  \\ \hline
            \kw{DBLP}    & 13,107 & 4,335 & 10,413& 35  & 11,198 & 127 & 255 \\ \hline
            \kw{IMDB}    & 17,871 & 3,914 &9,736 & 40  & 13,689 & 177 & 183 \\ \hline
      \end{tabular}
      }
  }
  \vspace{-4mm}
\end{table}

\begin{table}[t]
  \centering
  \caption{Index size in GB (Exp. 2)}
  \vspace{-4mm}
  \label{tab:index-size}
  {\small
  \setlength{\tabcolsep}{2pt}
  \resizebox{\linewidth}{!}{
        \begin{tabular}{|l|r|r|r|r|r|r|r|}
        \hline
            \textbf{Dataset} & \textbf{\Surf} & \textbf{\kw{ComSURF}} & \textbf{\VectorMaton} & \textbf{\ACORN-1} & \textbf{\ACORN-$\gamma$} & \kw{HNSW} & \textbf{\kw{OneSparse}}\\ \hline \hline
                \SIFT        & 17.70 & 5.33 &15.05  & 0.91 & 1.40 & 0.73 & 2.12 \\ \hline
        \GIST        & 21.61 & 7.48 &18.38 & 4.01 & 4.67 & 3.71 & 4.80\\ \hline
        \Redcaps     & 27.98 & 7.96 & 24.60& 2.34 & 2.82 & 2.05 & 2.80\\ \hline
                \kw{Deep}    & 18.83 & 6.79 & 15.56 & 1.39 & 1.87 & 1.20 & 2.63  \\ \hline
        \arXiv       & 8.75  & 1.48 &4.98  & 0.52 & 0.73 & 0.47 & 0.76\\ \hline
        \kw{Laion}   & 41.64 & 8.43 &28.80 & 2.34 & 2.88 & 2.05 & 3.76 \\ \hline
                \kw{DBLP}    & 65.89 & 12.76& 38.79 & 3.29 & 3.78 & 3.12 & 6.20 \\ \hline
                \kw{IMDB}    & 53.12 & 11.84 & 21.09 & 4.25 & 4.74 & 4.07 & 6.36\\ \hline
      \end{tabular}
      }
  }
  \vspace{-4mm}
\end{table}

\noindent\textbf{Exp.~2: Indexing Cost.}
Tables~\ref{tab:build-time} and~\ref{tab:index-size} report the index construction time (in seconds) and index size (in GB), respectively.
\Milvus, \VBASE, and Post-filtering rely on the same underlying HNSW index; therefore, we report their indexing costs using the HNSW results.
Compared with the state-of-the-art \VectorMaton, \kw{ComSURF} achieves $2.28$--$5.70\times$ faster index construction and $1.78$--$3.42\times$ smaller index size across all eight datasets.
Compared with the original \Surf, \kw{ComSURF} further reduces construction time by $2.23$--$6.43\times$ and index size by $2.77$--$5.91\times$, demonstrating the effectiveness of our compression rules.
Together with the superior search performance observed in Exp.~1, these results demonstrate that \kw{ComSURF} achieves both efficient index construction and query processing.
While \kw{ACORN} and \kw{OneSparse} require smaller index space and construction time, their search performance is substantially worse than \Surf/\kw{ComSURF}, as shown in Exp.~1.

\begin{figure}[t]
  \centering
  \includegraphics[width=0.95\linewidth]{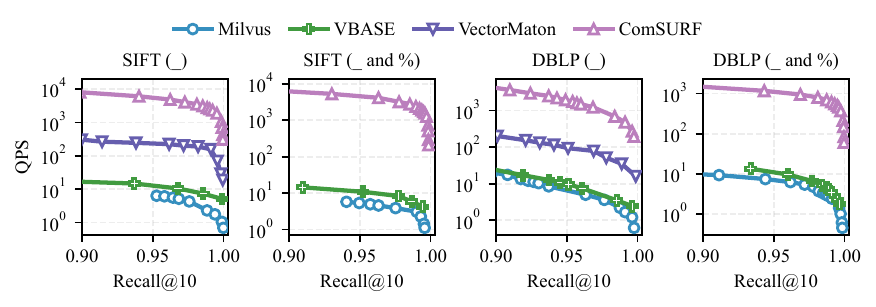}
  \vspace{-4mm}
  \caption{Search performance with \texttt{LIKE} predicates (Exp. 3)}
  \label{fig:like-qps-recall}
  \vspace{-3mm}
\end{figure}

\noindent\textbf{Exp.~3: Search Performance with \kw{LIKE} Predicates.}
We evaluate vector searches on \SIFT\ and \kw{DBLP} using two query workloads: (1) patterns with only \texttt{\_}, and (2) patterns with both \texttt{\_} and \texttt{\%}. Queries are generated by injecting wildcards into substring predicates while keeping the selectivity at approximately 0.2. We compare \kw{ComSURF} against two native \texttt{LIKE} baselines, \Milvus~\cite{WangYGJXLWGLXYY21} and \VBASE~\cite{ZhangXCSXCCH00Y23}, and \VectorMaton, whose wildcard support is limited to \texttt{\_}. We set $k=10$ and report the search performance in Fig.~\ref{fig:like-qps-recall}. Since the indexes from Exp.~2 are reused, we omit index costs.

Across both datasets, \kw{ComSURF} outperforms \Milvus\ and \VBASE\ by two to three orders of magnitude in QPS at comparable recall. For the \texttt{\_}-only workload, \kw{ComSURF} also achieves over an order of magnitude higher QPS than \VectorMaton. This advantage stems from our link-tree-based pruning, which avoids exploring redundant branches induced by wildcards. In contrast, \VectorMaton explores all outgoing automaton transitions upon encountering \texttt{\_}, incurring additional overhead. Meanwhile, \Milvus\ and \VBASE\ evaluate \texttt{LIKE} predicates during ANN search on a global index, resulting in costly traversal and predicate evaluation over non-qualifying vectors.

\begin{figure}[t]
  \centering

  \subfigure[Average update time]{
    \includegraphics[width=0.35\linewidth]{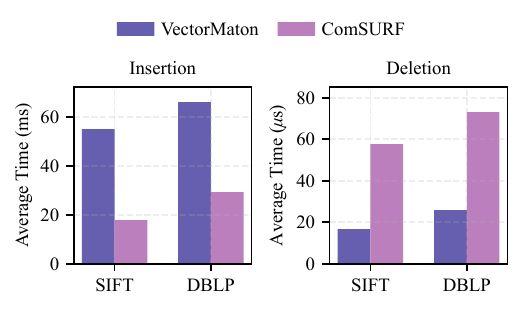}
    \label{fig:update-time}
  }
  \subfigure[Search performance before and after updates]{
    \includegraphics[width=0.58\linewidth]{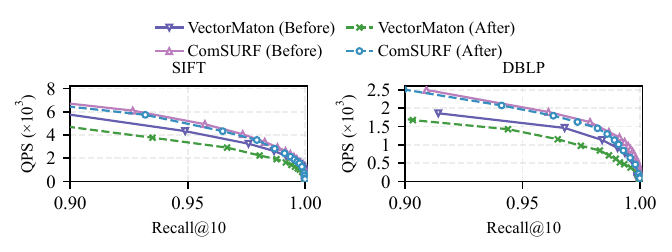}
    \label{fig:search-maintain}
  }

  \vspace{-4mm}
  \caption{Index maintenance performance (Exp. 4)}
  \label{fig:maintain}
  \vspace{-3mm}
\end{figure}

\noindent\textbf{Exp. 4: Index Maintenance.}
Following~\cite{abs-2603-01525}, we first build an index on 80\% of the data, then insert the remaining 20\%, followed by deleting 20\% of the indexed data. We report the average insertion and deletion times in Fig.~\ref{fig:update-time}, and compare the search performance before and after index maintenance in Fig.~\ref{fig:search-maintain}.
\kw{ComSURF} achieves faster insertion than \VectorMaton, benefiting from our efficient index construction techniques.
However, its deletion is slower due to our periodic index 
reconstruction strategy, whereas \VectorMaton adopts only lazy deletion.
Despite this additional maintenance cost, \kw{ComSURF} largely preserves its search performance after updates, while \VectorMaton exhibits noticeable performance degradation.

\begin{table}[t]
  \centering
  \caption{The effects of $\alpha$ and $ef_{\min}$ on index cost (Exp. 5)}
  \vspace{-4mm}
  \label{tab:param-cost}
  {\small
  \setlength{\tabcolsep}{3pt}
  \resizebox{\linewidth}{!}{
  \begin{tabular}{|r||r|r|r|r|r|r|r|r|}
    \hline
    $\alpha$ & 0.3&0.3&0.3&0.3&0.1&0.5&0.7&1\\ \hline
    $ef_{\min}$ & 0&100&200&500&200&200&200&200\\ \hline
    SIFT Index Size (GB) & 7.67 &5.55 &5.33 &5.05 &4.23 &5.76 &5.99 &6.20 \\ \hline
    SIFT Build Time (s) & 519 &499 &416 &408 &389 &429 &438 &457  \\ \hline
    DBLP Index Size (GB) & 16.72 &13.16 &12.76 &12.23 &10.55 &14.03 &14.57 &15.42 \\ \hline
    DBLP Build Time (s) & 4,564 &4,351 &4,335 &4,025 &3,291 &4,738 &4,960 &5,422 \\ \hline
  \end{tabular}
  }
  }
  \vspace{-4mm}
\end{table}

\begin{figure}[t]
  \centering
  \includegraphics[width=0.95\linewidth]{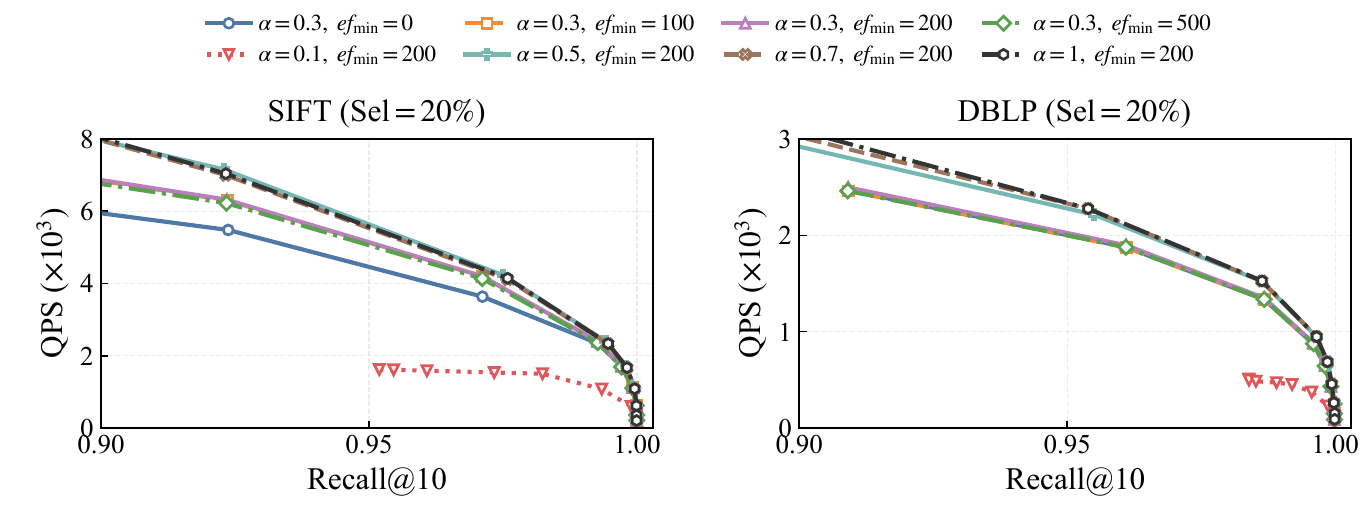}
  \vspace{-5mm}
  \caption{The effects of $\alpha$ and $ef_{\min}$ on search (Exp. 5)}
  \label{fig:qps-recall-param-sel-0.2}
  \vspace{-4mm}
\end{figure}

\begin{figure}[!t]
  \centering
  \begin{minipage}{0.48\linewidth}
    \includegraphics[width=\linewidth]{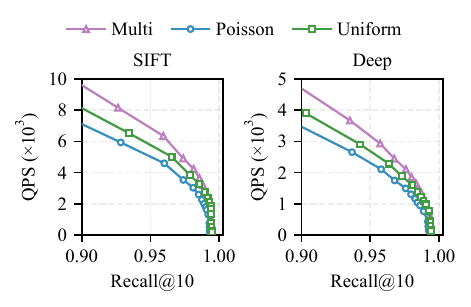}
  \vspace{-4mm}
  \caption{Varying string attribute distributions (Exp. 6)}
  \label{fig:qps-recall-distribution}
  \end{minipage}
  \hspace{1mm}
  \begin{minipage}{0.48\linewidth}
    \includegraphics[width=\linewidth]{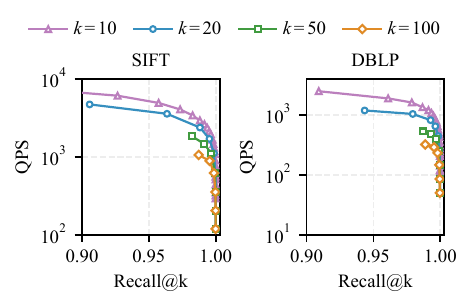}
  \vspace{-4mm}
  \caption{Varying obtained number of results $k$ (Exp. 7)}
  \label{fig:qps-recall-k}
  \end{minipage}
  
\end{figure}

\noindent\textbf{Exp.~5: Effects of Parameters $\alpha$ and $ef_{\min}$.}
We evaluate the effects of $\alpha$ and $ef_{\min}$ on search performance and index cost. Setting $ef_{\min}=0$ disables prefiltering, while $\alpha=1$ restricts ancestor serving to
states with identical candidate-set sizes. As shown in Table~\ref{tab:param-cost}, increasing $ef_{\min}$ reduces index size and building time, but may affect search performance. On \SIFT, search performance slightly degrades when $ef_{\min}$ increases from 200 to 500, whereas on DBLP, it remains nearly unchanged. This suggests that dataset-specific tuning of $ef_{\min}$ can further improve the tradeoff between search performance and index cost. For $\alpha$, larger values generally improve search performance by reducing postfiltering overhead, at the expense of higher index size and construction time. These results demonstrate the potential benefits of parameter tuning, which we leave for future work to automate.

\noindent\textbf{Exp.~6: Varying Synthetic Attribute Distributions.}
We study the impact of synthetic string distributions on performance using \SIFT and \kw{Deep} as examples, with query selectivity fixed at $0.2$.
We additionally generate base-string attributes under three distributions: Multinomial, Poisson, and Uniform, while generating query predicates using the same procedure described in \textit{Datasets}.
Fig.~\ref{fig:qps-recall-distribution} reports the QPS-$recall@10$ curves.
\kw{ComSURF} performance is stable across different distributions, and skewed distributions, Zipf and Multinomial, are slightly easier for all methods because frequently repeated substrings induce denser filtered neighbors.

\noindent\textbf{Exp.~7: Vary $k$.}
We evaluate the impact of $k \in \{10, 20, 50, 100\}$ on the search performance of \kw{ComSURF}, with selectivity fixed at 0.2.
As shown in Fig.~\ref{fig:qps-recall-k}, search performance gradually decreases as $k$ increases on both \SIFT\ and \kw{DBLP}, since retrieving more nearest neighbors requires examining and maintaining more candidates to achieve comparable recall.
Nevertheless, the performance gap becomes smaller at high recall levels.

\section{Conclusion}
\label{sec:con}

In this paper, we propose \Surf/\kw{ComSURF} for substring filtered ANN search.
We first design \kw{MultiSA}, a provably minimal suffix automaton for organizing substring-equivalent object sets, and integrate it with proximity graphs to avoid redundant traversal over non-qualifying objects.
We further develop efficient index compression techniques and extend our approach to support \kw{LIKE} predicates with two wildcards.
Extensive experiments on 8 real-world datasets demonstrate the effectiveness and efficiency of our approach.
\kw{ComSURF} consistently achieves superior search performance over existing baselines across different query selectivities.
Compared with the state-of-the-art approach, \VectorMaton, \kw{ComSURF} achieves $1.78$--$3.42\times$ smaller index size and $2.28$--$5.70\times$ faster index building, while delivering better search performance.

\balance
\bibliographystyle{ACM-Reference-Format}
\bibliography{reference}

\numberwithin{theorem}{section}
\numberwithin{example}{section}

\appendix

\setcounter{section}{0}
\renewcommand{\thesection}{\Alph{section}}
\renewcommand{\thesubsection}{\thesection.\arabic{subsection}}

\numberwithin{theorem}{section}
\numberwithin{lemma}{section}

\end{document}